# A Core-Valence Separated Similarity Transformed EOM-CCSD Method for Core-excitation Spectra

Santosh Ranga, and Achintya Kumar Dutta*

*Department of Chemistry, Indian Institute of Technology Bombay, Powai, Mumbai 400076.*

We present the theory and implementation of a core-valence separated similarity transformed EOM-CCSD (STEOM-CCSD) method for K-edge core excitation spectra. The method can select an appropriate active space using CIS natural orbitals and near 'black box' to use. The second similarity transformation Hamiltonian is diagonalized in the space of single excitation. Therefore, the final diagonalization step is free from the convergence problem arising because of the coupling of the core-excited states with the continuum of doubly excited states. Convergence trouble can appear for the preceding core-ionized states calculation in STEOM-CCSD. A core-valence separation scheme (CVS) compatible with the natural orbital based active space selection has been implemented to overcome the problem. The CVS-STEOM-CCSD has similar accuracy as that of the standard CVS-EOM-CCSD method but comes with a lower computational cost. The modification required for the CVS scheme because of the use of CIS natural orbital is highlighted. The suitability of CVS-STEOM-CCSD for chemical application is demonstrated by simulating the K-edge spectra of glycine and thymine.

*achintya@chem.iitb.ac.in

## I. Introduction:

The use of core-electron spectroscopy to study properties of matter has a long history[1]. The highly localized nature of the core-orbitals and the sizeable difference of energy between the core-orbitals of different elements make the core-excitation spectroscopy a tool for atom specific probe of the electronic structure of the material. However, the high-energy X-ray radiations required to excite the core-electron have hindered the widespread use of core electron spectroscopy. In recent years, considerable advances have been made in the quality and availability of X-ray sources[2]; even tabletop instruments have arrived. The X-ray absorption spectroscopy (XAS) is now routinely used to elucidate the structure, composition, and electronic distribution of molecules, surfaces and materials[3].

The proper analysis and interpretation of experimental results often require an in-depth understanding of the theory behind the phenomenon. Theoretical simulations are now routinely used for interpreting experimentally measured X-ray spectra. Various theoretical methods for x-ray spectroscopy simulation, starting from semi-empirical[4,5], DFT[6–15] to the state-of-the-art wave-function-based methods[16–21], are described in the literature. The readers are requested to consult ref[22] and ref[23] for the recent updates in the field. Among the various single-reference methods available in the literature for the theoretical simulations, the coupled cluster method[24] has emerged as the most accurate and systematically improvable one. The coupled-cluster method can be extended to excited states to excited states using the equation of motion (EOM-CC) approach[25–29]. The EOM-CC is generally used in singles and doubles approximation (EOM-CCSD), which scales as O ($N^6$) for both ground and excited states. The equation of motion coupled cluster approach is extremely successful in simulating excited state dominated by single excitation and is particularly successful[30] for valence and Rydberg excited states. However, the challenges in the simulation of core excited states can differ significantly from that observed in the valence excited states. A large relaxation effect is associated with the excitation of tightly bound core-electron, which is difficult to treat adequately using the linear excitation operator in EOM, especially when working with the singles and doubles truncation of the excitation operator.

Nooijen and Bartlett[31] have used the open-shell electron attached equation of motion coupled cluster method (OS-EA-EOMCC) to compute the core-excitation energy of small molecules. The OS-EA-EOMCC method or its SAC-CI analog (OS-EA-SAC-CI)[32] separates the correlation effect from the relaxation effect and predicts core-excitation energies, which are in good agreement with the experiments. However, both methods have the inherent problem of solving the coupled cluster method for core-ionized reference state, which sometimes induce severe convergence problems. A similar problem can also occur in the maximum overlap based coupled cluster[33] method for core-excitation spectra. The standard EOM-CCSD method has been found out to systematically overestimated the core excitation energies due to the missing relaxation effect. However, one can use a constant shift to properly align the simulated spectra with the experimental one. The recent implementation of core-valence separation[34] in EOM-CC by Coriani and co-workers[35,36] has made EOM-CCSD an efficient option for the simulation of core excitation spectra. Numerous studies[18,35–43] has been reported describing the application of CVS-EOM-CCSD method for core-ionized and core-excited states for K and L-edge spectra. Extension to non-linear properties like RIXS cross-section has also been achieved[44,45]. Bartlett and co-workers have reported a time-dependent EOM-CCSD implementation for the core-excitation spectra[46]. However, the $O(N^6)$ scaling and the associated have restricted the application of EOM-CCSD beyond small molecules, especially for the XAS, where many roots need to be calculated for the simulation of experimental spectra. Various strategies such as perturbative approximation[47–51], use of semi-numerical approximations[52,53], density fitting[54,55], Cholesky decomposition[54], approximations based on local and natural orbitals have been described in the literature[56–65] to reduce the scaling of the EOM-CCSD methods.

Nooijen and co-workers have employed an alternative strategy of using a second similarity transformation to decouple the single excitation space from the higher-order excitations[66,67]. This strategy leads to a significant reduction in computational cost. The second similarity transformed EOM-CCSD (STEOM-CCSD) has emerged as an attractive option for extending the coupled cluster method to the excited states of large molecules[64,68,69]. In recent years, we have seen a surge in the new developments within the framework of STEOM-CCSD. Efficient implement[64], lower scaling approximations[62], automatic active space selection scheme[70], the extension to the vibrational problem[71], open-shell systems[72], and spin-orbit coupling[73] has been reported. One of the nice features of STEOM-CCSD is that it has uniform accuracy[64] for valence, Rydberg and charge-transfer excited states when the electron excitation happens from the valence occupied orbitals. Moreover, the increase in computation cost of STEOM-CCSD with the number of roots scales only $O(N^5)$ times the number of roots, as opposed to $O(N^6)$ times the number of roots for EOM-CCSD. Therefore, STEOM-CCSD can be advantageous over EOM-CCSD for the simulation of XAS, where a large number of states needs to be calculated for the simulation of the experimental X-ray spectra. However, the application of the STEOM-CCSD method for core-level spectroscopy requires implementation of core-valence separation which is compatible with the CIS natural orbital based active space selection scheme. This paper aims to extend the STEOM-CCSD method for core-excited states.

The paper is organized as follows. The next section gives the theory and computational details of the calculation.—Section III presents the benchmark of the accuracy of STEOM-CCSD for core-excited states and application of it for the simulation and interpretation of experimental K-edge spectra. The conclusion and future plans are presented in section IV.

## II. Theory and Computational Details:

### A. STEOM-CCSD

The main idea of STEOM-CCSD is based on two-fold many-body similarity transformation[74] of the Hamiltonian such that the most important one- and two-body net-excitation operators in the resulting transformed Hamiltonian vanish. It is followed by diagonalization of the second similarity transformed Hamiltonian in the singles subspace. One assumes that a single Hartree–Fock (HF) single determinant ($|\phi_0\rangle$) provides an accurate zeroth-order description of the ground state. The occupied orbitals in the HF determinant are denoted by indices i, j, k, and l, while virtual orbitals are denoted by a, b, c and d.

The first transformation is identical to that in EOMCC and is given by

$$\hat{\bar{H}} = e^{-\hat{T}} H e^{\hat{T}} \tag{1}$$

The operator $\hat{T}$ in the singles and doubles approximation

$$\hat{T} = \hat{T}_1 + \hat{T}_2 = \sum_{i,a} t_i^a \left\{ \hat{a}^\dagger \hat{i} \right\} + \frac{1}{4} \sum_{i,j,a,b} t_{ij}^{ab} \left\{ \hat{a}^\dagger \hat{i} \hat{b}^\dagger \hat{j} \right\} \tag{2}$$

Where curly braces indicate normal ordering with respect to $|\phi_0\rangle$ and Einstein summation convention has been used throughout the study.

The singly transformed Hamiltonian in equation (1) can be expressed in normal ordered, second quantized notation as

$$\hat{\bar{H}} = h_0 + \sum_{p,q} h_{pq} \left\{ \hat{p}^{\dagger}\hat{q} \right\} + \frac{1}{4} \sum_{p,q,r,s} h_{pqrs} \left\{ \hat{p}^{\dagger}\hat{r}\hat{q}^{\dagger}\hat{s} \right\} \tag{3}$$

The determining equations for $\hat{T}$ are obtained by setting one- and two-body pure excitation operators in $\hat{\bar{H}}$ to zero.

$$h_{ai} = \langle \phi_i^a | \hat{\bar{H}} | \phi_0 \rangle = \langle \phi_i^a | e^{-\hat{T}} \hat{H} e^{\hat{T}} | \phi_0 \rangle = 0 \tag{4}$$

$$h_{abij} = \langle \phi_{ij}^{ab} | \hat{\bar{H}} | \phi_0 \rangle = \langle \phi_{ij}^{ab} | e^{-\hat{T}} \hat{H} e^{\hat{T}} | \phi_0 \rangle = 0 \tag{5}$$

Equations (4) and (5) are essentially standard CCSD amplitude equations. The constant term $h_0$ in the normal ordered Hamiltonian gives coupled cluster energy.

$$h_0 = \langle \phi_0 | \hat{\bar{H}} | \phi_0 \rangle = \langle \phi_0 | e^{-\hat{T}} \hat{H} e^{\hat{T}} | \phi_0 \rangle = E_{cc} \tag{6}$$

In the standard EOMCC method, the singly transformed Hamiltonian is diagonalized over a suitable set of configurations to obtain ionized, attached, or excited states of the reference state. In STEOMCC, one performs a second similarity transformation.

$$\hat{G} = \left\{ e^{\hat{S}} \right\}^{-1} \hat{\bar{H}} \left\{ e^{\hat{S}} \right\} \tag{7}$$

The transformation operator $\hat{S}$ in singles and doubles truncation is defined as

$$\hat{S} = \hat{S}^{IP} + \hat{S}^{EA} \tag{8}$$

$$\hat{S}^{IP} = \sum_{i',m} S_{i'}^{m} \left\{ \hat{m}^{\dagger}\hat{i}' \right\} + \frac{1}{2} \sum_{i,m,b,j} S_{ij}^{mb} \left\{ \hat{m}^{\dagger}\hat{i}\hat{b}^{\dagger}\hat{j} \right\} \tag{9}$$

$$\hat{S}^{EA} = \sum_{a',e} S_{e}^{a'} \left\{ \hat{a}'^{\dagger} e \right\} + \sum_{a,b,e,j} S_{ej}^{ab} \left\{ \hat{a}^{\dagger}\hat{e}\hat{b}^{\dagger}\hat{j} \right\} \tag{10}$$

In the above equations, m and e denote active indices of the hole and particle type, respectively, while a prime denotes a restriction to orbitals that are not active.

Now, attention needs to be devoted to equation (7). As the components of $\hat{S}$ do not commute, a normal ordered exponential is used to simplify the details of the equations[66]. The inverse of the normal ordered exponential operator may not always be well defined. Therefore, equation (7) is generally expressed in an iterative fashion.

$$\hat{G} = \hat{\bar{H}}\left\{e^{\hat{S}}\right\} - \left\{e^{\hat{S}} - 1\right\}\hat{G} \tag{11}$$

Now one can use a connected form of the above quantity

$$\hat{G} = \left(\hat{\bar{H}}\left\{e^{\hat{S}}\right\}\right)_c - \left(\left\{e^{\hat{S}} - 1\right\}\hat{G}\right)_c \tag{12}$$

where subscript c denotes the connectedness of the species. The second similarity transformed Hamiltonian can be represented in the second quantized normal ordered form as

$$\hat{G} = g_0 + \sum_{p,q} g_q^p \left\{\hat{p}^\dagger \hat{q}\right\} + \frac{1}{4}\sum_{p,q,r,s} g_{rs}^{pq}\left\{\hat{p}^\dagger \hat{r} \hat{q}^\dagger \hat{s}\right\} + ....... \tag{13}$$

The amplitudes of the operator $\hat{S}$ are defined in such a way that matrix elements of the transformed Hamiltonian in second quantized notation become equal to zero

$$\bar{\bar{g}}_{i'}^{m} = \bar{\bar{g}}_{ij}^{mb} = \bar{\bar{g}}_{e}^{a'} = \bar{\bar{g}}_{ej}^{ab} = 0 \tag{14}$$

In addition, the zeros which pre-existed in $\bar{H}$ after solving the CCSD equations remain preserved.

$$\bar{\bar{g}}_{i}^{a} = \bar{\bar{g}}_{ij}^{ab} = 0 \tag{15}$$

and

$$\bar{\bar{g}}_{o} = h_0 = E_{cc} \tag{16}$$

The structure of the doubly transformed Hamiltonian in the N-particle space is

$$\begin{pmatrix} G & 0 & S & D & T \\ 0 & g_0 & X & X & X \\ S & 0 & X & X & X \\ D & 0 & \sim & X & X \\ T & \sim & \sim & \sim & X \end{pmatrix} \tag{17}$$

where 0, S, D, T represent the reference state, and singly, doubly, and triply excited determinants, respectively. The ~ in the above equation indicates matrix elements, which are very small in magnitude. These small matrix elements arise due to remaining two-body terms in $\hat{G}$ that involve inactive orbitals and three-body and higher-body matrix elements introduced by the transformations. If the ~ matrix elements vanish, the total matrix assumes a block diagonal structure, and the eigenvalues of the entire matrix can be found from the diagonalization of the individual sub-blocks.

However, the above-mentioned condition is only a 'good approximation'. Moreover, the transformation with $\{e^{S_1}\}$ does not change the particle-rank of an operator. Consequently, the final matrix in STEOMCCSD is diagonalized is

$$\hat{G} = \{e^{\hat{S}_2}\}^{-1} \bar{H} \{e^{\hat{S}_2}\} \tag{18}$$

and it provides access to both singlet and triplet excited states that are dominated by singly excited configurations.

The STEOM-CC is closely related to Fock-space multi-reference coupled cluster method[75–79]. The $\widehat{S_2}$ can be obtained by solving equation (14), which is practically the same as solving Fock-space multi-reference coupled cluster equations for (1,0) and (0,1) sectors. However, solving equation (14) can lead to convergence issues due to the intruder state problems. Therefore, the $S_2$ amplitudes are obtained by putting intermediate normalizations on the converged IP and EA-EOM-CCSD eigenvectors.

$$S_{ji}^{bm} = \sum_{\lambda} r_{ji}^{b}(\lambda)\, r_{\lambda m}^{-1}$$

$$S_{je}^{ba} = -\sum_{\lambda} r_j^{ba}(\lambda)\, r_{\lambda e}^{-1}$$

## B. Active space selection and core-valence separation for STEOM-CCSD

The excitation energy obtained from the diagonalization of $\hat{G}$ depends on the no of active orbital m and e chosen for the calculation. Here one should note that the active space in STEOM-CCSD is quite different from the active space in CAS based methods and has nothing to do with non-dynamic correlation. The active space in STEOM-CCSD denotes the dynamic correlation due to the quasi-particle. Nevertheless, one needs to choose an active space, and an appropriately chosen active space is essential for obtaining accurate excitation energy in STEOM-CCSD.

We have used an automatic active space selection scheme based on CIS natural orbital to calculate core-excitation energy in STEOM-CCSD. The active-space selection scheme is similar in spirit to that used for valence excitation[70]. However, it requires significant modifications to be compatible with the core-valence separation required for K-edge spectra. The singly excited/ionized states from the core orbital in the EOM-CCSD method generally remains embedded in the continuum of doubly excited/ionized states, which leads to convergence issues in the Davidson iterative diagonalization step. One possible solution is to project out such doubly excited states from the excited/ionized state wave-function, which will decouple the singly ionized/excited states from the space of double excitation. However, in STEOM-CCSD, the final $\hat{G}$ matrix is diagonalized in the singles space, and the doubly excited states are automatically decoupled from the single excitation space because of the second similarity transformation. Consequently, the iterative diagonalization of the second similarity transformed Hamiltonian does not suffer from the convergence problem as observed for EOM-CCSD. However, the preceding IP calculation for the generation of $\hat{S}_2$ amplitudes can suffer convergence problems due to the single ionization from the core-orbital being embedded in the continuum of the so-called satellite states.

Consequently, one needs to use the core-valence separation for IP calculation, and the CIS natural orbital based automatic active space selection scheme needs to be modified accordingly. The description of active space selection and the corresponding core-valence separation is provided below. For all the equations in the rest of the paper, $I, J...$ denotes occupied core-orbital $i_v, j_v...$valence occupied orbitals and $i, j...$ denotes any arbitrary occupied orbital (both core and valence).

After the solution of the RHF equation, a partial integral transformation is performed to generate the Coulomb and exchange type integrals, and the CIS problem is solved. The CIS guess vectors are generated in such a that it can excite an electron only from the designated core-orbitals. Subsequently, the roots whose eigenvector has maximum overlap with the guess vectors are followed during the Davidson iterative diagonalization procedure. After the solution of the CIS equation, we calculate the occupied and virtual block of the state-averaged CIS 1-particle density

$$D_{ij} = \frac{1}{N_\lambda} \sum_{\lambda,a} R_a^i(\lambda) R_a^i(\lambda) \quad (19)$$

$$D_{ab} = \frac{1}{N_\lambda} \sum_{\lambda,i} R_a^i(\lambda) R_b^j(\lambda) \quad (20)$$

Where $R_i^a$ is a CIS eigenvector corresponding to the core-excited state $I$. The total number of states used for the state averaging (N) is the same as the number of core-excited states one is interested in. The state-average CIS natural orbitals are obtained by diagonalizing the occupied and virtual block of the density matrix.

Occupied and virtual orbitals up to a certain threshold are considered active, and in ORCA, the thresholds are controlled by the keyword IPTHRESH and EATHRESH, respectively.

The default threshold truncation parameter has been chosen to be 0.001 after extensive testing.

Next, the Fock matrix is transformed into the natural orbital basis

$$F' = U^{\dagger} F U \quad (21)$$

The active and inactive blocks of both occupied(o) and virtual subspace(v) is are then separately diagonalized to get a block diagonal Fock matrix.

$$F'' = V^{\dagger} F' V = \begin{pmatrix} \varepsilon^{o}_{i_v} & F''_{i_v I} & 0 & 0 \\ F''_{I i_v} & \varepsilon^{o}_{I} & 0 & 0 \\ 0 & 0 & \varepsilon^{v}_{A} & F''_{A\tilde{A}} \\ 0 & 0 & F''_{\tilde{A}A} & \varepsilon^{v}_{\tilde{A}} \end{pmatrix} \quad (22)$$

In the above matrix $e$ represents diagonal blocks of the Fock matrix, $F''_{Ii_v}, F''_{i_v I}, F''_{A\tilde{A}}, F''_{\tilde{A}A}$, represents non-zero off diagonals blocks of the Fock matrix, and $A$, $\tilde{A}$ represents active and inactive virtuals, respectively. The Brillouin condition is valid for $F''$ and

$$F''_{i_v A} = F''_{i_v \tilde{A}} = F''_{IA} = F''_{I\tilde{A}} = F''_{A i_v} = F''_{\tilde{A} i_v} = F''_{AI} = F''_{\tilde{A}I} = 0 \quad (23)$$

In the next step, the Hartree-Fock AO to MO coefficients are updated as

$$C' = C U^{\dagger} V^{\dagger} \quad (24)$$

Where $C'$ is the transforms matrix from AO to CIS NO basis. The CIS vectors are also transformed to NO basis and saved to be used as a guess in the Davidson diagonalization of $\hat{G}$ matrix.

$$R'(l) = R(l) U^{\dagger} V^{\dagger} \quad (25)$$

In the next step, AO integral is transformed into the CIS NO basis, and subsequent steps in STEOM-CCSD is performed in the CIS NO basis only.

One interesting thing to notice that in equation (22) highest occupied orbitals are core orbital from which the excitation is considered. Therefore, one can directly solve the IP-EOM-CCSD equations corresponding to the ionization from the core orbital without even have to solve for the valence IP. However, the solution IP-EOM-CCSD equation for core ionization in CIS NO basis can face the convergence issues similar to that observed in the canonical MO basis. One needs to use the core-valence separation for the IP-EOM-CCSD step in STEOM-CCSD. However, due to the difference in the structure of the Fock Matrix, the structure of the Hamiltonian in CVS-IP-EOM-CCSD will differ in a canonical and CIS NO basis. In the case of canonical molecular orbitals, the highest occupied molecular orbitals are of valence type. On the other hand, the highest occupied molecular orbitals are of core-type in the CIS NO basis. Figure 1 gives a schematic representation of the CVS-IP-EOM-CCSD Hamiltonian in canonical MO and CIS natural orbital basis. The programmable expressions for the CVS-IP-EOM-CCSD sigma equations are presented in Appendix I.

In the original CVS-EOM-CCSD scheme of Coriani and co-workers[35], the ground state $\hat{T}$ amplitudes are solved using frozen core approximation

$$h_{ai_v} = \langle f_{i_v}^{a} | e^{-\hat{T}} \hat{H} e^{\hat{T}} | f_0 \rangle = 0$$

$$h_{abi_v j_v} = \langle \phi_{i_v j_v}^{ab} | e^{-\hat{T}} \hat{H} e^{\hat{T}} | \phi_0 \rangle = 0$$

With $t_I^a = 0, t_{IJ}^{ab} = 0$ and $t_{i_v J}^{ab} = 0$. This can reduce the computational cost of the ground state coupled cluster calculations. However, $F_{i_v}^{I}$ is non-zero in the CIS natural orbitals basis, and therefore, we did not freeze the core orbitals during the solution of the ground state coupled cluster amplitudes. One can avoid this problem by doing the transformation to the CIS natural orbital basis after the solution of the ground state coupled cluster, which will allow one to use frozen core approximation for the ground state coupled cluster amplitudes. However, such an approach will require an additional integral transformation step and is not considered in the present study.

The modified expression for the $\hat{S}_{IP}$ amplitudes in CVS-STEOM-CCSD will be as follows

$$S_{jI}^{ma} = -\sum_{\lambda} r_{jI}^{a}(\lambda) r_{\lambda m}^{-1}$$

$$S_{Ji_v}^{ma} = -\sum_{\lambda} r_{Ji_v}^{a}(\lambda) r_{\lambda m}^{-1}$$

The expression for $\hat{S}_{EA}$ will be the same as the valence case. The modified expressions for the $\hat{G}$ intermediates are presented in appendix II.

The STEOM-CCSD transition moments are calculated using EOM-CCSD like expectation value approach, as described in ref [80]. The modified expression for the doubles part of excited state right and left and right eigenvectors are for CVS-STEOM-CCSD as follows

$$\tilde{r}_{ab}^{ij} = \frac{1}{\sqrt{2}}(1 + P_{ij}P_{ab})\left(S_{ej}^{ab} r_i^e - S_{iJ}^{mb} r_m^a - S_{Ij_v}^{mb} r_m^a\right)$$

$$\tilde{l}_{ab}^{ij} = \frac{1}{\sqrt{2}}(1 + P_{ij}P_{ab})\left(S_{ej}^{ab} l_i^e - S_{iJ}^{mb} l_m^a - S_{Ij_v}^{mb} l_m^a\right)$$

STEOM-CCSD natural transition orbitals are calculated using a CIS like approximation using $\tilde{l}_i^a$ and $\tilde{r}_i^a$ vectors. The core-valence separated STEOM-CCSD (denoted as CVS-STEOM-CCSD in the rest of the paper), as described above, is implemented in the development version of the quantum chemistry program package ORCA[81]. It should be noted that all the perturbative, local, and natural orbital based, density fitting and semi-numerical approximation defined for standard STEOM-CCSD can be used CVS-STEOM-CCSD with minor modifications. However, in this paper, we keep our attention focused on standard STEOM-CCSD with only CVS approximation.

## III. Result and discussion:

**A Benchmarking for small molecules:**

To benchmark the newly implemented CVS-STEOM-CCSD method's accuracy, we have used the small molecule test set of Mathews[82] for which EOM-CCSDT results have been reported. The test set consists of five molecules $H_2O$, CO, $NH_3$, HCN, and $C_2H_4$, with a total of 28 core excited states. Experimental results are available for some of the excited states.

Table 1 reports CVS-STEOM-CCSD values for K-edge core-excited states in the aug-cc-pCVQZ basis. The excitation energies are heavily overestimated (around 2 eV) as compared to the EOM-CCSDT or experimental results. The trend is similar to that observed for the EOM-CCSD[82] method. The excitation of the core electron accompanies a large orbital-relaxation effect, which cannot be adequately included in the linear excitation operator in EOM-CCSD. The STEOM-CCSD method uses an exponential operator ($\left\{e^{\hat{S}}\right\}$) for the excited state. However, the use of the normal order ansatz[83] prevents contractions between the $\hat{S}$ operators. It results in linear $\hat{S}$ amplitudes in the IP and EA equations and cannot bring in the high orbital relaxation effect for the core-excitation. This leads to a blue shift of the simulated XAS as compared to experimental spectra. However, it can still reproduce the peak separation to a good extend, and proper alignment with the experimental spectra can be achieved by applying a constant shift to the simulated spectra. Moreover, one can accurately reproduce the experimental term values which are of high interest as they are characteristic of bonding features in the molecule. The term values can be calculated as the difference of the core-excitation energy ($w_{EE}$) and core-ionization energy (the core-excitation energy ($w_{IP}$)). The mean absolute error in the terms values in the CVS-STEOM-CCSD is 0.23 eV with respect to EOM-CCSDT results (See Table 2). The CVS-EOM-CCSD shows a similar error bar of 0.21 for the same benchmark set, and the accuracy of terms values in CVS-STEOM-CCSD are very similar to the CVS-EOM-CCSD method (see Figure 2(a)).

Compared to the experimental term values, both STEOM-CCSD and EOM-CCSDT methods show quite similar performance (see Figure 2(b)). It has been shown[84] before for core-ionization that even EOM-CCSDT cannot give quantitative accuracy, and one needs to go for quadrupoles for a proper agreement with the experimental results. We believe the same principles also apply to core excitations. However, one should not draw any far-reaching conclusion about the relative accuracy of STEOM-CCSD and EOM-CCSD from the limited comparison performed in this study.

Table 3 presents the trend in CVS-STEOM-CCSD K edge excitation energy values with respect to the increase in the basis set. We have used the hierarchy of aug-cc-pCVXZ family of basis sets, with X=D, T, and Q. It can be seen that the excitation energy is particularly sensitive to the size of the basis set. The core excitation energy shows a considerable change (in the range of 1-2 eV) on going from the aug-cc-pCVDZ basis set to the aug-cc-pCVTZ basis set. The change from the aug-cc-pCVTZ basis to the aug-cc-pCVQZ is much smaller and is less than 0.3 eV for all the excited states in the Mathews test set. The creation of core-hole results in a large perturbation of the reference state wave-function. One needs to use a flexible basis set with a large number of basis functions to capture the large orbital relaxation effect observed for core-ionized/core-excited state. Coriani and co-workers[82] have recently shown that the use of the partially or fully uncontracted basis can lead to quicker convergence with respect to the size of the basis set.

It is also essential to understand the effect of basis set on simulated spectra, as one can always align the lowest energy peak with the experimental peak using a constant shift. Figure 3 presents the simulated Oxygen K-edge spectrum for $N_2O$ in the aug-cc-pCVXZ (X=D, T, and Q) basis set along with the experimental spectra. It can be seen that going from the aug-cc-pCVDZ to the aug-cc-pVTZ basis can result in better agreement with the experimental spectra. However, the qualitative trends in the experimental spectra can still be predicted with aug-cc-pCVDZ level simulated spectra. The spectrum appears to be nearly identical in the aug-cc-pCVTZ and aug-cc-pCVQZ basis set, when a constant shift is applied to align with the lowest energy peak.

One needs to investigate the effect of the CVS approximation in the STEOM-CCSD calculations. The core-excitation energy in STEOM-CCSD can be calculated without the core-valence separation in a smaller aug-cc-pCVDZ basis set. Table 4 presents the K-edge excitation energy and the term value in STEOM-CCSD with and without the CVS approximation. The error in the excitation energy and term values due to CVS approximation is within 0.01 eV, which is outside the error bar of the STEOM-CCSD method itself. To investigate the suitability of CVS-STEOM-CCSD for the experimental spectrum simulation, we have investigated the K-edge spectrum of glycine and thymine.

### B. Glycine K edge Spectra

Glycine is the smallest among all the amino acids. It exists as a neutral molecule in the gas phase and as a zwitterion in the aqueous solvent[85,86]. A large number of X-ray spectroscopy studies in the gas phase[87,88] and the solvation phase[89] have been reported in the literature. It makes Glycine an ideal model system for the preliminary chemical application of our CVS-STEOM-CCSD method. The same system was used by Coriani and co-workers to investigate the performance of their linear response coupled-cluster method[38]. Due to the internal hydrogen bonding, Glycine has a large number of stable conformers. To compare with the experimental K-edge spectra, we have calculated the core-excitation energy of the lowest energy conformer of glycine (See Figure 4). The geometry of Glycine is taken from ref 38. The atoms are numbered in the ascending order of EOM-CCSD IP values. All the subsequent CVS-STEOM-CCSD calculations were done using the aug-cc-pCVDZ basis set with additional diffuse 5s5p4d functions added to the center of mass of Glycine. The diffuse functions were added in an even tapered way where the successive exponents were generated by multiplying the previous exponent with a factor of 0.33. The additional diffuse functions and all the excitation energies were presented in the SI. The experimental spectra of Glycine are taken from ref[87] .

Figure 5 presents the experimental and simulated Carbon K-edge spectra of Glycine. The simulated spectrum is shifted by -1.3 eV to align with the experimental spectra. The Carbon K-edge experimental spectra give a strong asymmetric peak, which arises due to transition from1s of (C=O) Carbon ($C_1$) to $p^*$ (C=O) (See Figure 6). The experimental spectra show a small feature B with band maxima at 289.4 eV. The peak arises due to the transition from the 1s of $C_\alpha$ ($C_2$) to $p^*$ (C-O). A weak, broad peak(C) appears in the range 290 eV to 293 eV, which arises due to the transitions from the 1s of $C_\alpha$ . The most dominant among them are two transitions. One is $C_a \rightarrow p^*\left(C - N\right)$, which also have some Rydberg character, and the other is a Rydberg type of transition from the 1s of $C_\alpha$ (See Figure 6).

The O K edge spectra of Glycine is presented in Figure 7. The simulated spectrum is shifted by -3.1 eV to align with the experimental spectrum. The experimental Oxygen K-edge spectrum shows two sharp bands in the low energy region. As one can see from the NTOs in Figure 8, the first peak experimental peak at 532.2 eV (A) arises due to excitation from the 1s of carbonyl Oxygen (O4) to the $p^*$ (C=O). The second peak with band maxima at 535.4 eV arises due to the excitation of 1s of OH oxygen (O5). It can be resolved to the combination of two peaks, one is due to the transition from 1s of OH oxygen to the $p^*$ (C=O), and the other is due to the transition from 1s of OH oxygen to $s^*$ of OH oxygen. Next, in the experimental spectra, two broadband with band maxima at 537.7 eV and 539.2 eV. They arise primarily due to the Rydberg transitions mostly from OH oxygen.

The experimental and simulated Nitrogen K-edge spectra of Glycine is presented in Figure 9. The simulated spectrum is shifted by -3.04 eV to align with the experimental spectra. The experimental Nitrogen K-edge spectra of Glycine shows four prominent bands. The lowest two bands have a band maxima at 401.2 eV and 402.4 eV, respectively. Band A has a width of 0.5 eV and arises due to the transition from Nitrogen 1s to $\sigma^*$ (N-H) orbital. Band B has a width of 0.6 eV and is assigned to the $N1s \rightarrow \pi^*_{C-N}$ transition. The higher energy part of the experimental spectra consists of two broad bands with band maxima at 403.8 eV and 404.8 eV, respectively. The band C arises due to the transition from nitrogen 1s to $s^*$ (N-C). The fourth broad band D can be assigned to three main transitions, two of which is Rydberg in nature and the third one arises due to the transition to $p^*$ (C-N) type resonance state (see Figure 10).

**C. Thymine X-Ray absorption spectra:**

Nucleobases are the building block of the genetic materials and are among the most important biomolecule. A large number of theoretical[35,42,90] and experimental studies[91,92] are available for the K-edge spectra of nucleobases. The majority of the studies are performed in the gas phase. They allow one to analyze spectral profile without having to deal with the complexities arising due to the coupling with solvent molecules. We have taken thymine as a representative example (See Figure 11) to demonstrate the suitability of the CVS-STEOM-CCSD method for the simulation of K-edge spectra of chemically interesting molecules. The aug-cc-pCVDZ basis set with additional diffuse functions, similar to that used for the Glycine, is used for the simulations. The experimental spectra are taken from ref[91] .

Figure 12 presents the experimental and simulated Oxygen K-edge spectra of thymine. The simulated spectra are shifted by -3.7 eV to align with the experimental spectrum. The experimental spectra for Oxygen K-edge can be divided into two distinct parts. The low energy part consists of two intense bands A and B, with band maxima at 531.4 eV and 532.3 eV, respectively. They arise due to the transition from the 1s orbital of $O_1$ and $O_2$ to the $\pi^*$ orbital. Next in the experimental spectra is two broad bands C and D. The C band is located in the 535.1–536.1 eV interval, which has a contribution from the $O_1 1s \rightarrow p^*$ transition. The band C also has a contribution of the Rydberg transitions (with some mixing with $\pi^*$orbital) from both Oxygen. The band D is located at 536.7 eV to 537.2 eV near the ionization threshold of thymine Oxygen. The band is consists of mostly low-intensity Rydberg transition from 1s orbital of both the Oxygen. (see Figure 13).

Figure 14 present the experimental and simulated Nitrogen K-edge spectra of thymine. The peaks in theoretical spectra are shifted by -3.1 eV to align with the experimental spectra.

Our CVS-STEOM-CCSD calculation shows that the Nitrogen K-edge spectrum has a complex pattern with closely spaced transitions, even at the lower energy value. The lowest experimental peak A has a band maxima at 401.7 eV. It is predominantly due to two transitions $N_3 1s \rightarrow p^*$ and $N_4 1s \rightarrow p^*$. The second band maxima (B) at 402.6 eV, comprises four transitions. There are two near degenerate $N_3 1s \rightarrow p^*$ and $N_4 1s \rightarrow p^*$ transition where the latter has a three-time larger intensity than the former. The other two transitions are also near degenerate $N_3 1s \rightarrow \sigma^*_{N_3-H}$ and $N_4 1s \rightarrow \sigma^*_{N_4-H}$ with the former having slightly higher intensity. The spectral band C with band maxima with 404.1 eV arises primarily due to Rydber type transitions mixed with some contribution from the valence orbitals. The most dominant of them is a Rydberg transition from $N_3$1s orbital. The latter part of the band also has contributions from the Rydberg transition from $N_4$1s orbital and a valence type $N_4 1s \rightarrow \sigma^*_{N_4-H}$ transition. The next part of the experimental spectra consists of a broad band D located after 404.6 eV. It has a band maxim at 405.5 eV. The band maxima is just below the ionization threshold of thymine nitrogen at 406.7 eV. Several Rydberg and valence spectrum contribute to this band. The most dominant ones are shown in Figure 15.

Figure 16 presents the experimental and simulated Carbon K-edge spectra of thymine. The simulated spectrum is shifted by -2.3 eV to align with the experimental one. The lowest two bands A and B in the spectra at 284.9 and 285.9 eV, can be assigned to the transition from the 1s of both $C_5$ and $C_6$ to the $\pi^*$ orbital. The next spectral band consists of three features denoted by C, D, and E at 287.3 eV, 287.8 eV, and 288.4 eV. Feature C is well reproduced by our CVS-STEOM-CCSD calculations and can be assigned to the $C_7 1s \rightarrow \pi^*$ transition. The peak position at D and E are underestimated and overestimated, respectively, by CVS-STEOM-CCSD method. The feature at D can be assigned to $C_8 1s \rightarrow \pi^*$. The feature E presumably arising from the Ryberg transition from 1s of $C_7$ carbon with some possible contribution from the Rydber transition from the 1s of $C_5$ carbon. The most prominent absorption band in the spectrum is F, with its maximum at 289.4 eV. It derives its intensity mostly from the transition $C_9 1s \rightarrow \pi^*$ along with some small contribution from Rydberg type transition from 1s orbital of $C_7$ and $C_6$. These Rydberg transitions mainly contribute to the dense spectral structure at the low-energy side of the band F. It should be noted that the peak position due to transition $C_9 1s \rightarrow \pi^*$ in CVS-STEOM-CCSD is underestimated as compared to the experimental band maxima. The next region in the spectra is broadband G near the thymine carbon 1s ionization threshold. The feature arises mainly due $C_5 1s \rightarrow \pi^*$ mixed with some Rydberg character with additional contribution from the Rydberg transition from 1s orbital of $C_5$ and $C_7$ carbon.

## IV. Conclusions :

The paper describes the formulation and implementation of the core-valence separated STEOM-CCSD method for K-edge core-excitation spectra. The CVS-STEOM-CCSD can automatically select an appropriate active space using state average CIS natural orbitals. The difference between the CVS scheme in STEOM-CCSD with that used in the EOM-CCSD method is highlighted. The use of CVS approximation leads to a very negligible error for the STEOM-CCSDF K-edge core-excitation energy. The absolute values of the K-edge core-excitation energies in CVS-STEOM-CCSD are grossly overestimated as compared to the experimental results. The trend is similar to that obtained in the CVS-EOM-CCSD method. However, one can get a proper alignment of the CVS-STEOM-CCSD simulated spectra with the experimental one by applying a constant shift. The CVS-STEOM-CCSD gives a similar performance as that of the CVS-EOM-CCSD method but at a lower computation cost. The excitation energy values in CVS-STEOM-CCSD are extremely sensitive to the used basis set. One at least needs to use a triple zeta quality basis set to get a reasonable agreement with experimental results. The suitability of the CVS-STEOM-CCSD for chemical applications is demonstrated by simulating the experimental K-edge spectra of Glycine and Thymine.

The next obvious step will combine the CVS-STEOM-CCSD method with PNO based lower scaling approximations, which will allow one to calculate the K-edge spectra of medium size molecules routinely. Work is in progress towards that direction.

## Supporting Information

The Supporting Information is available

Cartesian coordinates of Glycine and Thymine, the CVS-STEOM-CCSD excitation energy values, and oscillator strengths are provided in the supporting information.

## Acknowledgment

The authors acknowledge the support from the IIT Bombay, IIT Bombay Seed Grant project, DST-Inspire Faculty Fellowship for financial support, IIT Bombay super computational facility, and C-DAC Supercomputing resources (PARAM Yuva-II, Param Bramha) for computational time.

Conflict of interest

The authors declare no competing financial interest.

## Appendix I:

$$\sigma^{I} = \underbrace{-\tilde{\tilde{F}}_{I}^{l} r^{l} + \tilde{F}_{d}^{l} \tilde{r}_{d}^{Il}}_{F} - \underbrace{\hat{g}_{Id}^{lM} \tilde{r}_{d}^{lM} - \hat{g}_{Id}^{Lm_v} \tilde{r}_{d}^{Lm_v}}_{3i}$$

$$\sigma_{b}^{iJ} = \underbrace{\tilde{\tilde{F}}_{d}^{b} r_{d}^{iJ} - \tilde{\tilde{F}}_{J}^{l} r_{b}^{il} - \overset{\approx}{F}{}_{i}^{l} r_{b}^{lJ} - t_{eb}^{iJ} \tilde{F}_{e}^{L} r^{L}}_{F} + \underbrace{\tilde{g}_{lM}^{iJ} \rho_{b}^{lM} + \tilde{g}_{Lm_v}^{iJ} \rho_{b}^{Lm_v}}_{4i} - \underbrace{\tilde{g}_{Lb}^{iJ} r^{L}}_{3i} + \underbrace{\tilde{\tilde{K}}_{bd}^{JL} r_{d}^{iL} + \tilde{\tilde{K}}_{bd}^{Jl_v} r_{d}^{Il_v} - \tilde{K}_{bd}^{JL} \rho_{d}^{Li} - \tilde{K}_{bd}^{Jl_v} \rho_{d}^{l_v I} - \tilde{J}_{db}^{il} \rho_{d}^{lJ}}_{2e}$$
$$- \underbrace{\tilde{r}_{e}^{mN} g_{fe}^{mN} t_{fb}^{iJ} - \tilde{r}_{e}^{Mn_v} g_{fe}^{Mn_v} t_{fb}^{iJ}}_{3b}$$

$$\sigma_{b}^{Ij_v} = \underbrace{\tilde{\tilde{F}}_{d}^{b} r_{d}^{Ij_v} - \tilde{\tilde{F}}_{j_v}^{l} r_{b}^{Il} - \overline{F}{}_{I}^{L} r_{b}^{Lj_v} - t_{eb}^{Ij_v} \tilde{F}_{e}^{L} r^{L}}_{F} + \underbrace{\tilde{g}_{Lm}^{Ij_v} \rho_{b}^{Lm} + \tilde{g}_{l_v M}^{Ij_v} \rho_{b}^{l_v M}}_{4i} - \underset{3i}{\tilde{g}_{Lb}^{Ij_v} r^{L}} + \underbrace{\tilde{\tilde{K}}_{bd}^{j_v l} r_{d}^{Il} - \tilde{K}_{bd}^{j_v l} \rho_{d}^{lI} - \tilde{J}_{db}^{IL} \rho_{d}^{Lj_v}}_{2e}$$
$$- \underbrace{\tilde{r}_{e}^{mN} g_{fe}^{mN} t_{fb}^{Ij} - \tilde{r}_{e}^{Mn_v} g_{fe}^{Mn_v} t_{fb}^{Ij_v}}_{3b}$$

Where

$$\rho_{b}^{iJ} = r_{b}^{iJ} + r^{I} t_{b}^{J}$$

$$r_{b}^{Ij_v} = r_{b}^{Ij_v} + r^{I} t_{b}^{j_v}$$

The $\tilde{F}$ ,$\tilde{\tilde{F}}$ , $\tilde{g}$ , $\hat{g}$ , $\tilde{K}$ , $\tilde{\tilde{K}}$ and $\tilde{J}$ are standard $\bar{H}$ intermediates and $g$ is the bare molecular integrals

## Appendix II:

The expressions for the G intermediates

$$\tilde{S}_{ij}^{mb} = 2S_{iJ}^{mb} + 2S_{Ij_v}^{mb} - S_{jI}^{mb} - S_{Ji_v}^{mb}$$

$$\tilde{S}_{ej}^{ab} = 2S_{ej}^{ab} - S_{ej}^{ba}$$

Ghh elements

$$u_m^i = \tilde{F}_k^b \tilde{S}_{ik}^{mb} - \hat{g}_{ib}^{kl} \tilde{S}_{kl}^{mb}$$

$$\bar{\bar{g}}_i^j = \tilde{\tilde{F}}_i^j + u_m^i \delta_{mj}$$

Gpp elements

$$u_{ae} = \tilde{F}_k^c \tilde{S}_{ek}^{ac} - \hat{g}_{cd}^{ld} \tilde{S}_{el}^{cd}$$

$$\bar{\bar{g}}_{ac} = \tilde{\tilde{F}}_c^a + \delta_{ce} u_{ac}$$

Gphph elements

$$\bar{\bar{g}}_{ab}^{ij}(J) = \tilde{J}_{ab}^{ij} + u_{ac}^{im} \delta_{mj} + u_{ae}^{ik} \delta_{eb} + u_{ae}^{im} \delta_{mj} \delta_{eb}$$

where

$$u_m^c = -g_{cd}^{kl} \tilde{S}_{kl}^{md}$$

$$u_k^e = g_{cd}^{kl} \tilde{S}_{el}^{cd}$$

$$u_{ac}^{im} = \tilde{F}_K^c S_{iK}^{ma} + \tilde{F}_{k_v}^c S_{ik_v}^{ma} + \hat{g}_{dc}^{la} \tilde{S}_{il}^{md} - \hat{g}_{cd}^{La} S_{iL}^{md} - \hat{g}_{cd}^{l_v a} S_{Il_v}^{md} + \hat{g}_{ic}^{kL} S_{kL}^{ma} + \hat{g}_{ic}^{Kl_v} S_{Kl_v}^{ma}$$

$$u_{ae}^{ik} = \tilde{F}_k^c S_{ei}^{ac} + \hat{g}_{id}^{kl} \tilde{S}_{el}^{ad} - \hat{g}_{id}^{lk} S_{el}^{ad} + \hat{g}_{cd}^{ka} S_{ei}^{cd}$$

$$u_{ae}^{im} = u_m^c S_{ei}^{ac} - u_e^K S_{iK}^{ma} - u_e^k S_{Ik_v}^{ma} + u_{il}^{mc} \tilde{S}_{el}^{ad} - u_{kd}^{im} S_{ek}^{ad} + \tilde{u}_{kd}^{im} S_{ek}^{da} + u_{ie}^{Kl} S_{lK}^{ma} + u_{ie}^{k_v L} S_{Lk_v}^{ma}$$

$$u_{il}^{mc} = g_{cd}^{kl} \tilde{S}_{mc}^{ik}$$

$$u_{kd}^{im} = g_{dc}^{kl} \tilde{S}_{mc}^{il}$$

$$\tilde{u}_{kd}^{im} = g_{dc}^{kL} S_{iL}^{mc} + g_{dc}^{kl_v} S_{Il_v}^{mc}$$

$$u_{ie}^{kl} = g_{cd}^{kl} S_{ei}^{cd}$$

Gphhp elements

$$\bar{\bar{g}}_{ab}^{ij}(K) = \tilde{K}_{ab}^{ij} + \tilde{\tilde{u}}_{ac}^{im} \delta_{mj} + \tilde{\tilde{u}}_{ae}^{ik} \delta_{eb} + \tilde{\tilde{u}}_{ae}^{im} \delta_{mj} \delta_{eb}$$

$$\tilde{\tilde{u}}_{ab}^{im} = \hat{g}_{ib}^{Kl} S_{lK}^{ma} + \hat{g}_{ib}^{k_v L} S_{Lk_v}^{ma} - \hat{g}_{ka}^{bc} S_{kI}^{mc} - \hat{g}_{Ka}^{bc} S_{Ki_v}^{mc} + \tilde{F}_k^b S_{kI}^{ma} + \tilde{F}_K^b S_{Ki_v}^{ma}$$

$$\tilde{\tilde{u}}_{ae}^{ij} = \bar{F}_k^e S_{ie}^{ac} + \hat{g}_{ja}^{cd} S_{ei}^{cd} - \hat{g}_{ic}^{kj} S_{ke}^{ac}$$

$$\tilde{\tilde{u}}_{ae}^{im} = u_m^c S_{ei}^{ca} - u_e^k S_{kI}^{ma} - u_e^K S_{Ki_v}^{na} + \tilde{\tilde{u}}_{kd}^{im} S_{ke}^{ad} + \tilde{u}_{ie}^{Kl} S_{lK}^{ma} + \tilde{u}_{ie}^{k_v L} S_{Lk_v}^{ma}$$

$$\tilde{\tilde{u}}_{kd}^{im} = g_{cd}^{kl} S_{lI}^{mc} + g_{cd}^{kL} S_{Li_v}^{mc}$$

$$\tilde{u}_{ie}^{kl} = g_{cd}^{kl} S_{ib}^{cd}$$

□

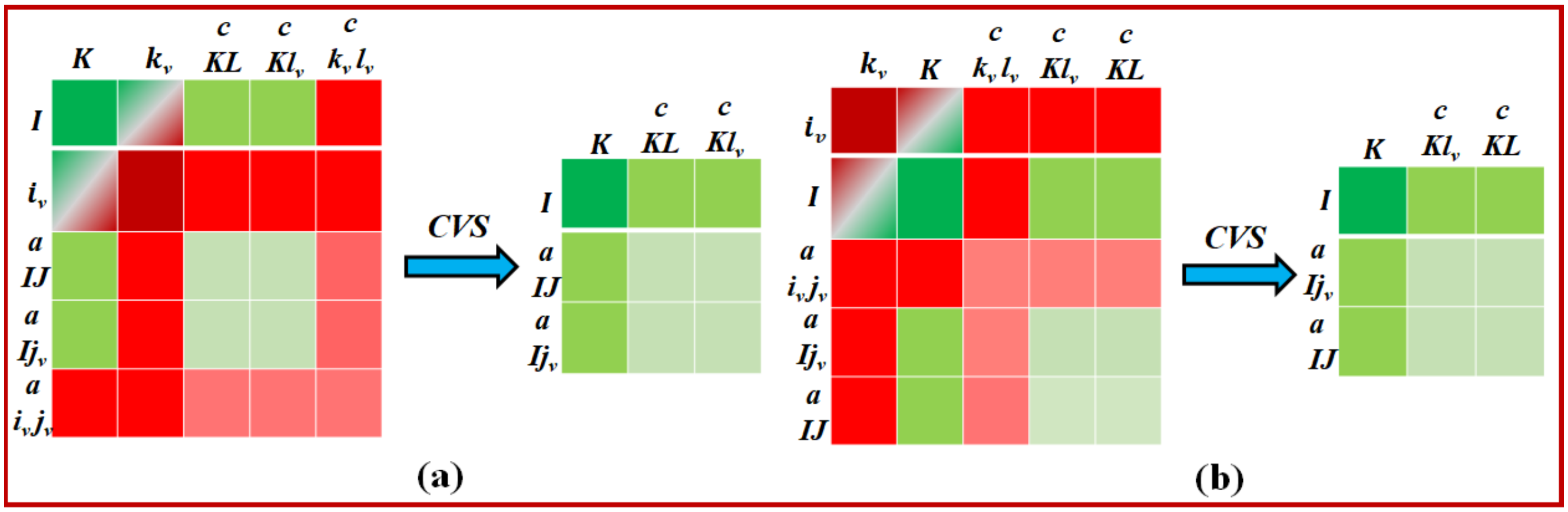

*Figure 1: The core-valence separation in the IP-EOM-CCSD step of (a) canonical (b) CIS natural orbital based STEOM-CCSD.*

***Table 1: Comparison of CVS-STEOM-CCSD K-edge values with the CVS-EOM-CCSD and CVS-EOM-CCSDT. The aug-cc-pCVQZ basis set used for the calculations.***

| **Molecule** | **Transition** | **CVS-EOM-CCSDT**[82] | **CVS-EOM-CCSD**[82] | **CVS-STEOM-CCSD** | **Exp** |
|---|---|---|---|---|---|
| | K→ π* | 285.081 | 286.103 | 286.075 | 284.7 |
| $C_2{*}H_4$ | K → 3s | 287.372 | 288.69 | 288.766 | 287.2 |
| | K → 3p | 288.014 | 289.33 | 289.439 | 287.9 |
| | K → 4p | 288.135 | 289.657 | 289.522 | -NA- |
| HCN* | K→ π* | 400.029 | 401.102 | 401.023 | 399.7 |
| | K → 3s | 402.746 | 404.783 | 404.321 | 402.5 |
| | K → 4s | 403.209 | 404.822 | 404.71 | -NA- |
| | K → 3p | 404.357 | 406.037 | 405.883 | -NA- |
| | K→ π* | 286.824 | 287.537 | 287.189 | 286.4 |
| HC*N | K → 3s | 289.526 | 290.422 | 290.155 | 289.1 |
| | K → 4s | 290.07 | 291.111 | 291.039 | 290.6 |
| | K → 3p | 291.158 | 292.14 | 291.991 | -NA- |
| | K → 3s | 533.96 | 535.593 | 535.767 | 534 |
| H2O* | K → 3p | 535.78 | 537.371 | 537.42 | 535.9 |
| | K → 4s | 537.3 | 539.03 | 539.149 | -NA- |
| | K → 4p | 537.39 | 539.06 | 539.204 | -NA- |
| | K → 3s | 400.764 | 402.129 | 402.14 | 400.8 |
| N*H3 | K → 3p | 402.437 | 403.789 | 403.8 | 402.5 |
| | K → 4s | 403.567 | 404.903 | 404.8 | 403.0 |
| | K → 5s | 404.317 | 405.796 | 405.465 | -NA- |
| | K→ π* | 287.664 | 288.182 | 288.044 | 287.3 |
| C*O | K → 3s | 292.868 | 294.038 | 294.072 | 292.5 |
| | K→ π*/3d | 293.924 | 295.081 | 294.991 | 293.4 |
| | K → 4s | 294.122 | 295.258 | 295.126 | -NA- |
| | K→ π* | 534.209 | 535.844 | 535.688 | 534.1 |
| CO* | K → 3s | 538.782 | 540.651 | 540.564 | 538.8 |
| | K→ π*/3d | 539.877 | 541.924 | 541.761 | 539.8 |
| | K → 4s | 539.969 | 542.222 | 541.909 | -NA- |

***Table 2: Comparison of term values with CVS-EOM-CCSD/CCSDT and STEOM method using aug-cc-pCVQZ basis set.***

| Molecule | Transition | CVS-EOM-CCSDT[82] | CVS-EOM-CCSD[82] | CVS-STEOM-CCSD | Exp |
|---|---|---|---|---|---|
| | K→ π* | -5.723 | -6.133 | -6.16 | -5.90 |
| **$C_2$*$H_4$** | K → 3s | -3.432 | -3.546 | -3.469 | -3.4 |
| | K → 3p | -2.79 | -2.906 | -2.796 | -2.7 |
| | K → 4p | -2.669 | -2.579 | -2.713 | NA |
| | K→ π* | -6.682 | -7.202 | -7.274 | -7.1 |
| **HCN*** | K → 3s | -3.965 | -3.521 | -3.976 | -4.3 |
| | K → 4s | -3.502 | -3.482 | -3.587 | NA |
| | K → 3p | -2.354 | -2.267 | -2.414 | NA |
| | K→ π* | -6.708 | -7.077 | -7.431 | -6.96 |
| **HC*N** | K → 3s | -4.006 | -4.192 | -4.465 | -4.26 |
| | K → 4s | -3.462 | -3.503 | -3.581 | -2.76 |
| | K → 3p | -2.374 | -2.474 | -2.629 | NA |
| | K → 3s | -5.504 | -5.81 | -5.639 | -5.9 |
| **$H_2$O*** | K → 3p | -3.684 | -4.032 | -3.986 | -4.0 |
| | K → 4s | -2.164 | -2.373 | -2.257 | -2.1 |
| | K → 4p | -2.074 | -2.343 | -1.68 | -1.5 |
| | K → 3s | -4.702 | -4.902 | -4.848 | -4.8 |
| **N*$H_3$** | K → 3p | -3.029 | -3.242 | -3.188 | -3.1 |
| | K → 4s | -1.899 | -2.128 | -2.18 | -2.6 |
| | K → 5s | -1.149 | -1.235 | -1.523 | -1.9 |
| | K→ π* | -8.767 | -9.438 | -9.57 | -8.8 |
| **C*O** | K → 3s | -3.563 | -3.582 | -3.542 | -3.6 |
| | K→ π*/3d | -2.507 | -2.539 | -2.623 | -2.7 |
| | K → 4s | -2.309 | -2.362 | -2.488 | NA |
| | K→ π* | -8.046 | -8.425 | -8.521 | -8.3 |
| **CO*** | K → 3s | -3.473 | -3.618 | -3.645 | -3.4 |
| | K→ π*/3d | -2.378 | -2.345 | -2.448 | -2.4 |
| | K → 4s | -2.286 | -2.047 | -2.3 | NA |

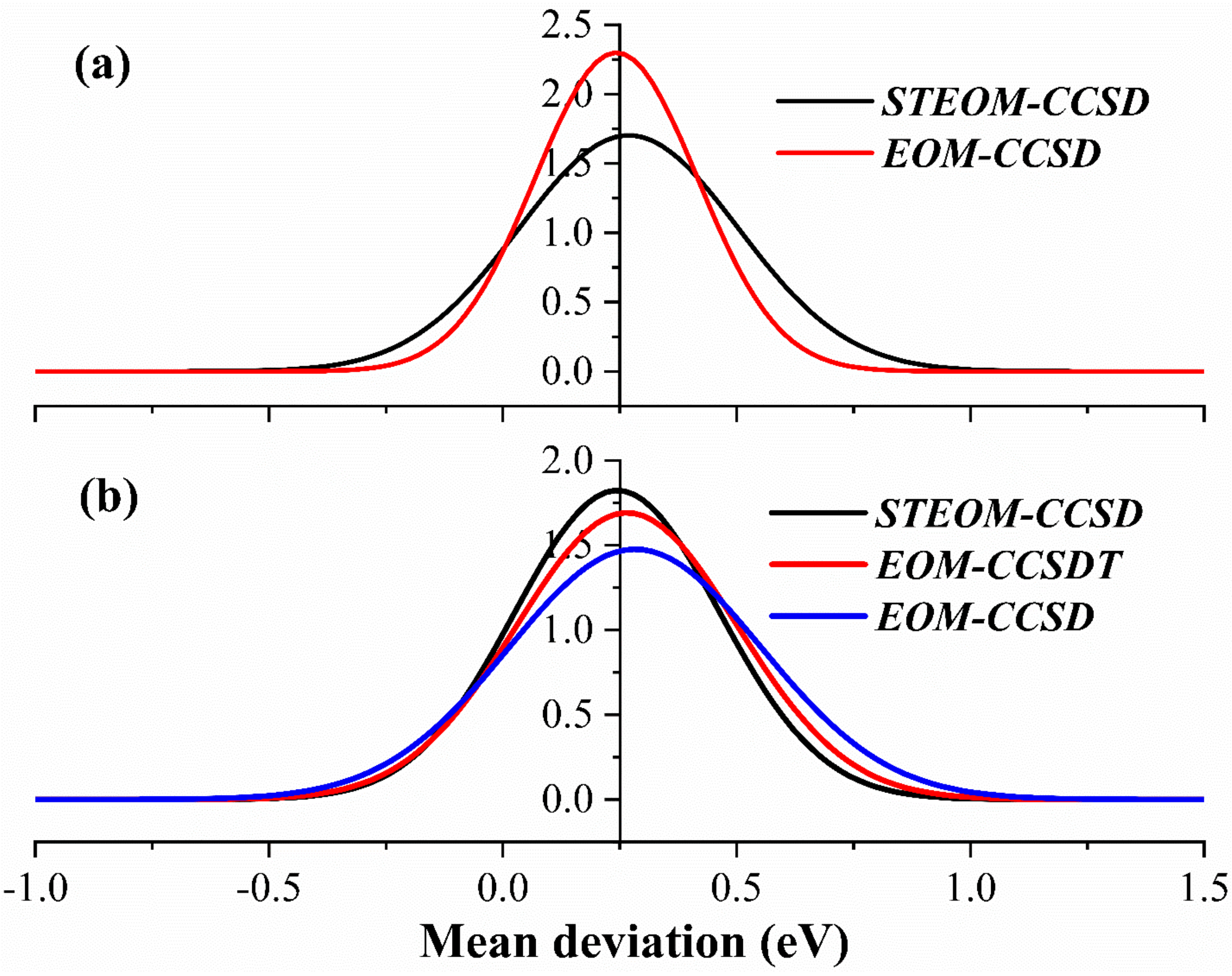

*Figure 2: Error distribution plot of CVS-STEOM-CCSD with respect to (a)EOM-CCSDT and (b) experimental results.*

**Table 3: The K-edge values with different basis sets using CVS-STEOM-CCSD method.**

| Molecule | Transition | aug-cc-pCVDZ | aug-cc-pCVTZ | aug-cc-pCVQZ |
|---|---|---|---|---|
| | K→ π* | 287.413 | 286.112 | 286.075 |
| $C_2$*$H_4$ | K → 3s | 289.904 | 288.782 | 288.766 |
| | K → 3p | 290.558 | 289.545 | 289.439 |
| | K → 4p | 290.680 | 290.746 | 289.522 |
| | K→ π* | 402.777 | 401.020 | 401.023 |
| HCN* | K → 3s | 406.028 | 404.531 | 404.321 |
| | K → 4s | 406.388 | 404.800 | 404.71 |
| | K → 3p | 407.784 | 406.053 | 405.883 |
| | K→ π* | 288.761 | 287.206 | 287.189 |
| HC*N | K → 3s | 291.529 | 290.201 | 290.155 |
| | K → 4s | 292.381 | 291.034 | 291.039 |
| | K → 3p | 293.493 | 292.105 | 291.991 |
| | K → 3s | 537.828 | 535.916 | 535.767 |
| $H_2$O* | K → 3p | 539.604 | 537.164 | 537.42 |
| | K → 4s | 541.751 | 539.409 | 539.149 |
| | K → 4p | 542.05 | 539.561 | 539.204 |
| | K → 3s | 403.663 | 402.173 | 402.14 |
| N*$H_3$ | K → 3p | 405.335 | 403.764 | 403.8 |
| | K → 4s | 406.775 | 405.028 | 404.8 |
| | K → 5s | 407.684 | 405.711 | 405.465 |
| | K→ π* | 289.28 | 287.736 | 288.044 |
| C*O | K → 3s | 295.381 | 294.068 | 294.072 |
| | K→ π*/3d | 296.403 | 295.057 | 294.991 |
| | K → 4s | 296.518 | 295.156 | 295.126 |
| | K→ π* | 538.021 | 535.81 | 535.688 |
| CO* | K → 3s | 542.742 | 540.692 | 540.564 |
| | K→ π*/3d | 544.026 | 541.913 | 541.761 |
| | K → 4s | 544.166 | 542.07 | 541.909 |

.

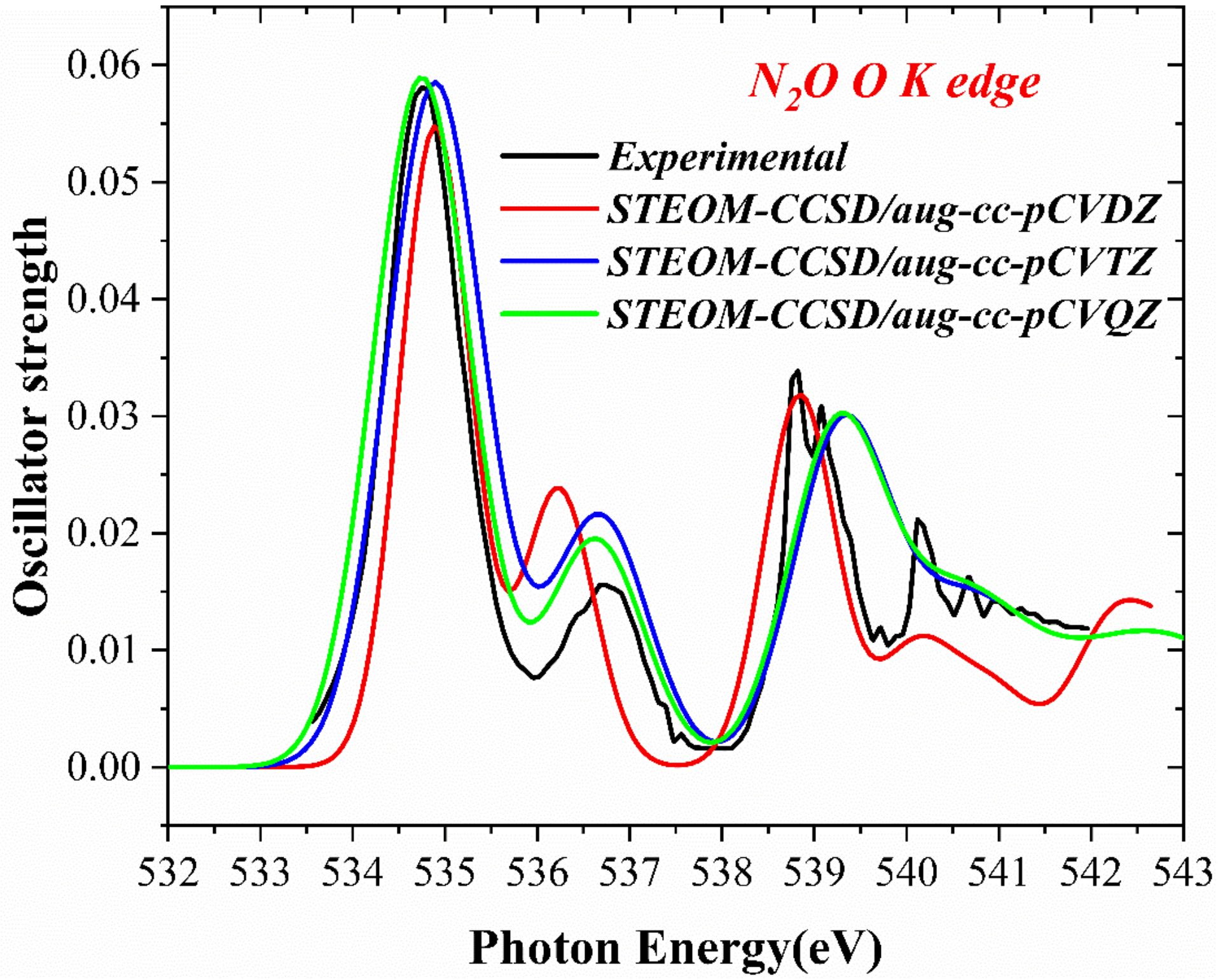

*Figure 3: Basis set dependence of the O K edge spectra of $N_2O$ molecule. The lowest energy peaks of all the simulated spectra is shifted to match the corresponding peak in the experimental spectra.*

***Table 4: Comparison of excitation energy and term values at STEOM-CCSD/aug-cc-pCVDZ theory: the effect of CVS***

| **Molecule** | **Transition** | **Excitation Energy** | | **Term Value** | |
|---|---|---|---|---|---|
| | | **CVS** | **no CVS** | **CVS** | **no CVS** |
| | K→ π* | 287.413 | 287.416 | -5.963 | -5.963 |
| $C_2^*H_4$ | K → 3s | 289.904 | 289.907 | -3.472 | -3.472 |
| | K → 3p | 290.558 | 290.560 | -2.816 | -2.819 |
| | K → 4p | 290.680 | 290.683 | -2.693 | -2.696 |
| | K→ π* | 402.777 | 402.767 | -7.073 | -7.08 |
| HCN* | K → 3s | 406.028 | 406.026 | -3.822 | -3.821 |
| | K → 4s | 406.388 | 406.384 | -3.462 | -3.463 |
| | K → 3p | 407.784 | 407.781 | -2.066 | -2.066 |
| | K→ π* | 288.761 | 288.757 | -7.141 | -7.146 |
| HC*N | K → 3s | 291.529 | 291.531 | -4.373 | -4.372 |
| | K → 4s | 292.381 | 292.381 | -3.521 | -3.522 |
| | K → 3p | 293.493 | 293.494 | -2.409 | -2.409 |
| | K → 3s | 537.828 | 537.822 | -5.626 | -5.626 |
| $H_2O^*$ | K → 3p | 539.604 | 539.597 | -3.849 | -3.851 |
| | K → 4s | 541.751 | 541.744 | -1.702 | -1.704 |
| | K → 4p | 542.05 | 542.043 | -1.403 | -1.405 |
| | K → 3s | 403.663 | 403.662 | -4.857 | -4.856 |
| $N^*H_3$ | K → 3p | 405.335 | 405.334 | -3.185 | -3.184 |
| | K → 4s | 406.775 | 406.773 | -1.745 | -1.744 |
| | K → 5s | 407.684 | 407.682 | -0.836 | -0.835 |
| C*O | K→ π* | 289.28 | 289.272 | -9.562 | -9.567 |
| | K → 3s | 295.381 | 295.376 | -3.461 | -3.463 |
| | K→ π*/3d | 296.403 | 296.398 | -2.439 | -2.441 |
| | K → 4s | 296.518 | 296.512 | -2.324 | -2.327 |
| | K→ π* | 538.021 | 538.011 | -8.187 | -8.19 |
| CO* | K → 3s | 542.742 | 542.734 | -3.466 | -3.467 |
| | K→ π*/3d | 544.026 | 544.018 | -2.182 | -2.183 |
| | K → 4s | 544.166 | 544.158 | -2.042 | -2.043 |

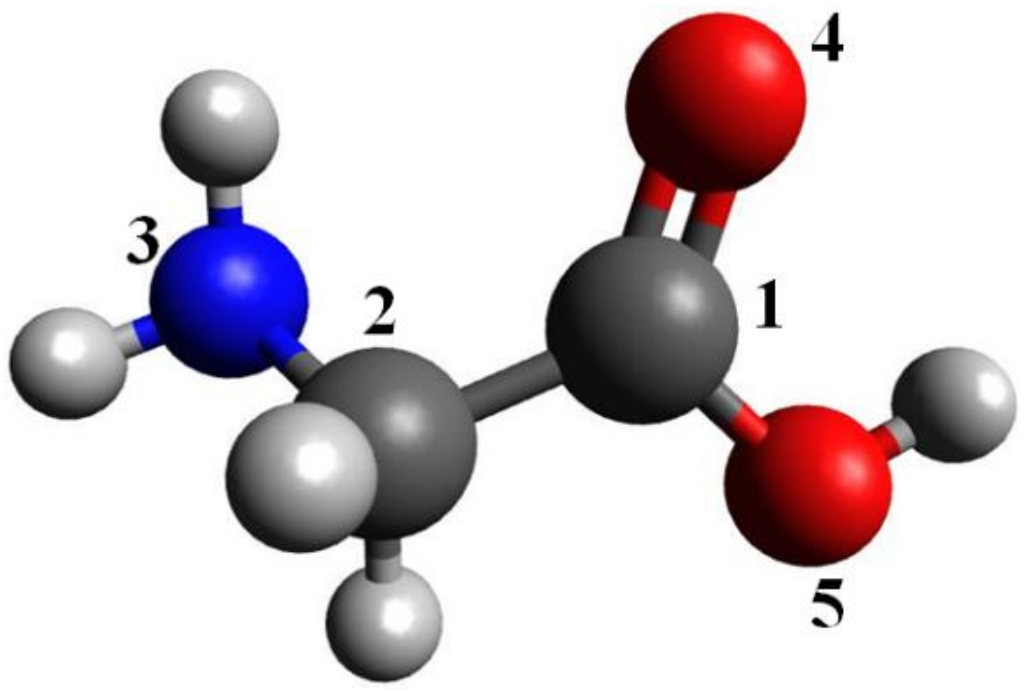

*Figure 4: Conformer I of Glycine taken from ref*[38] *in which the atoms are numbered according to ascending order EOM-CCSD IP value.*

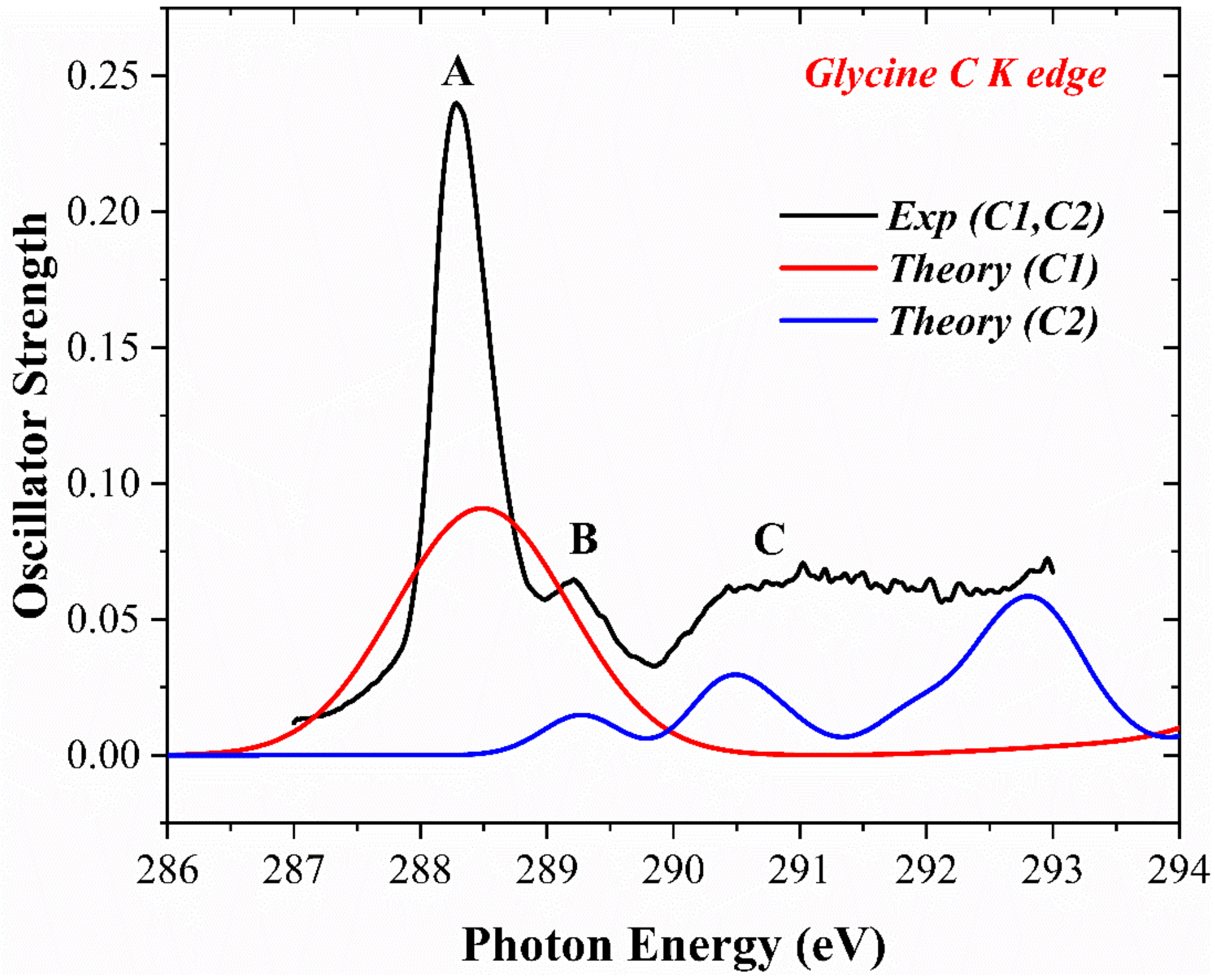

*Figure 5: Experimental and CVS-STEOM-CCSD simulated X-ray absorption spectra of C K-edge of Glycine. Simulated spectra were shifted by -1.3 eV.*

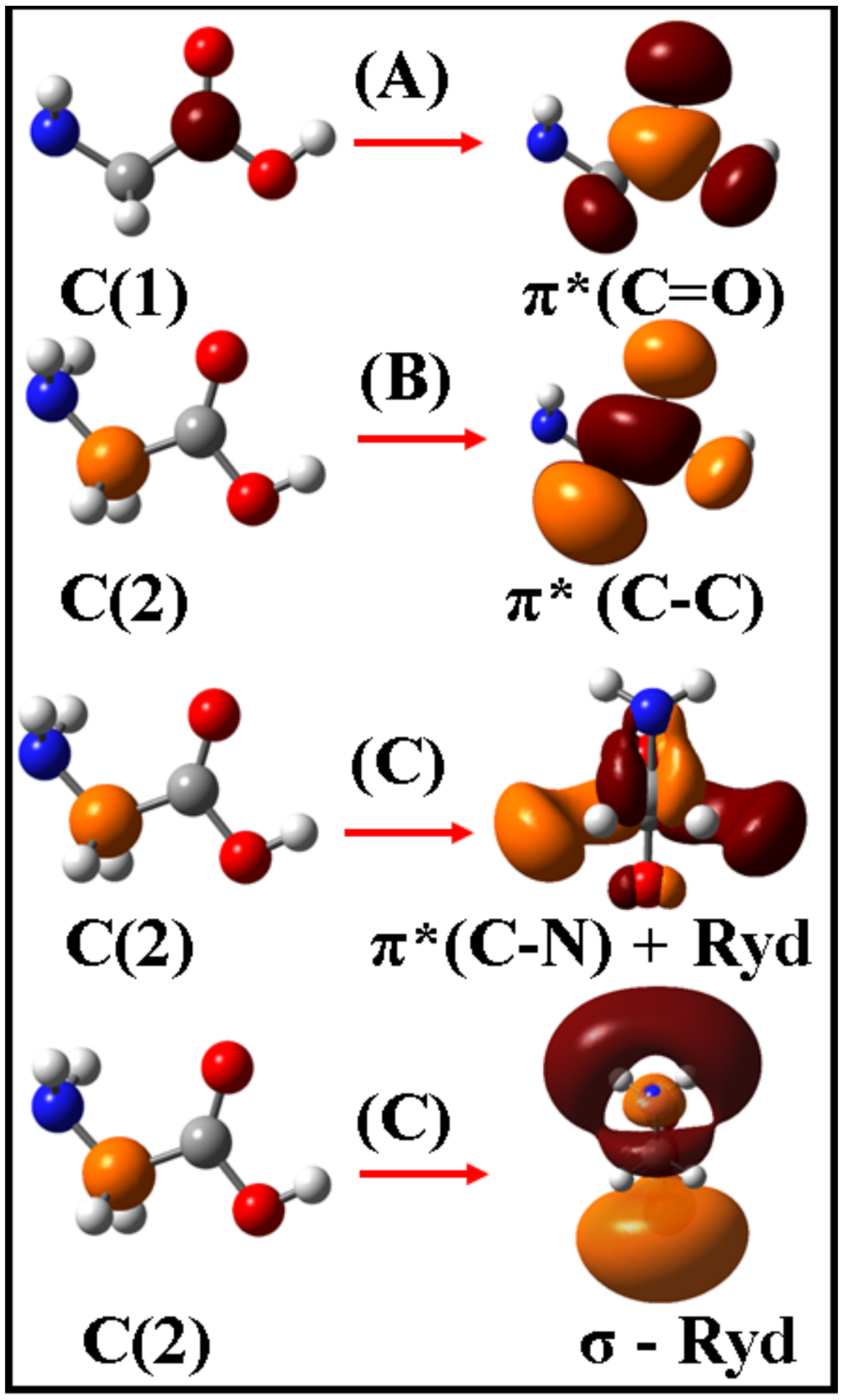

*Figure 6: Natural transition orbitals (NTOs) for C K-edge of glycine molecule.*

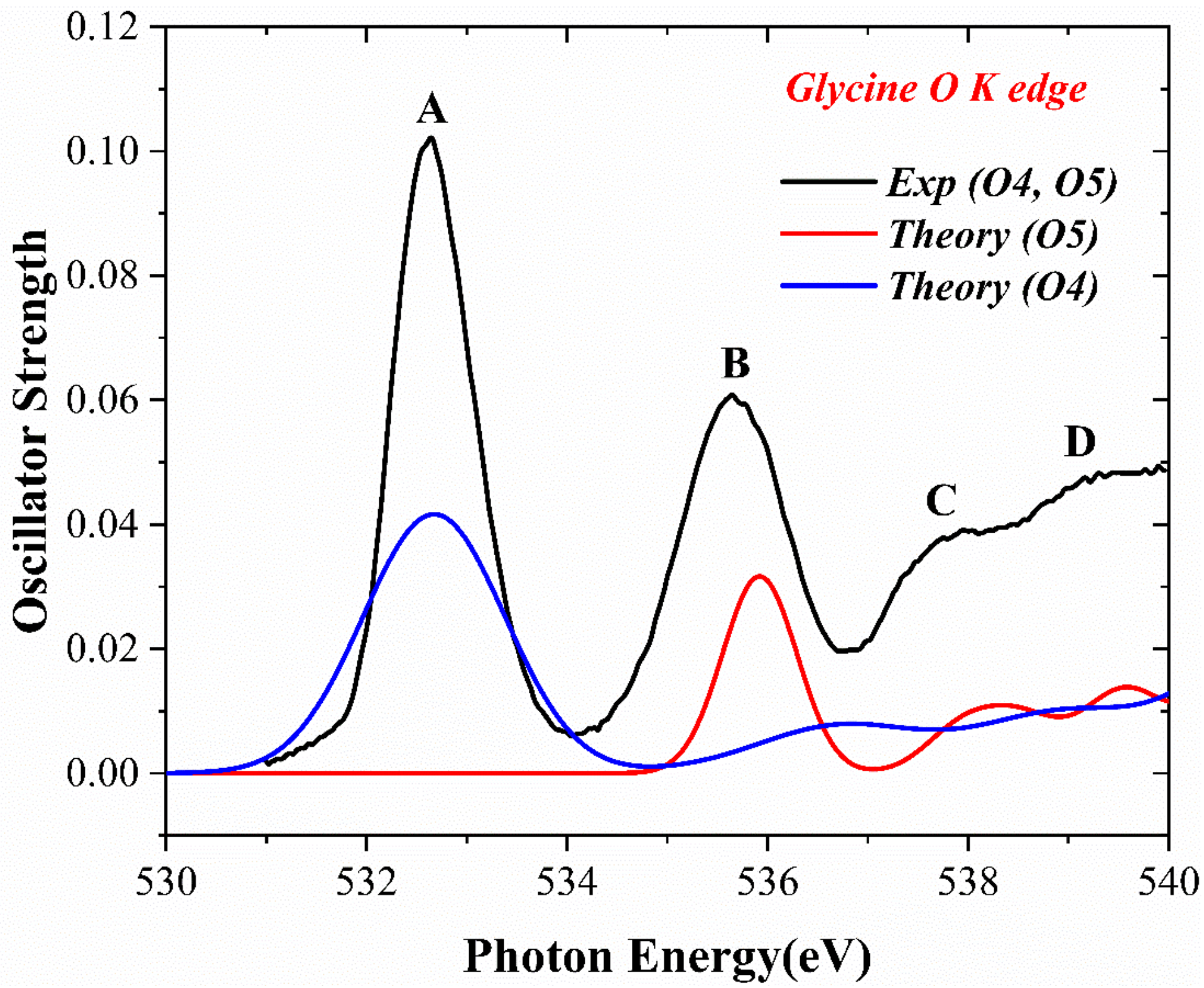

*Figure 7: Experimental and CVS-STEOM-CCSD simulated oxygen K edge spectra of Glycine using STEOM-CCSD/aug-cc-pCVDZ level of theory. The theoretical spectra are shifted by -3.3eV to match with the experimental peak separation.*

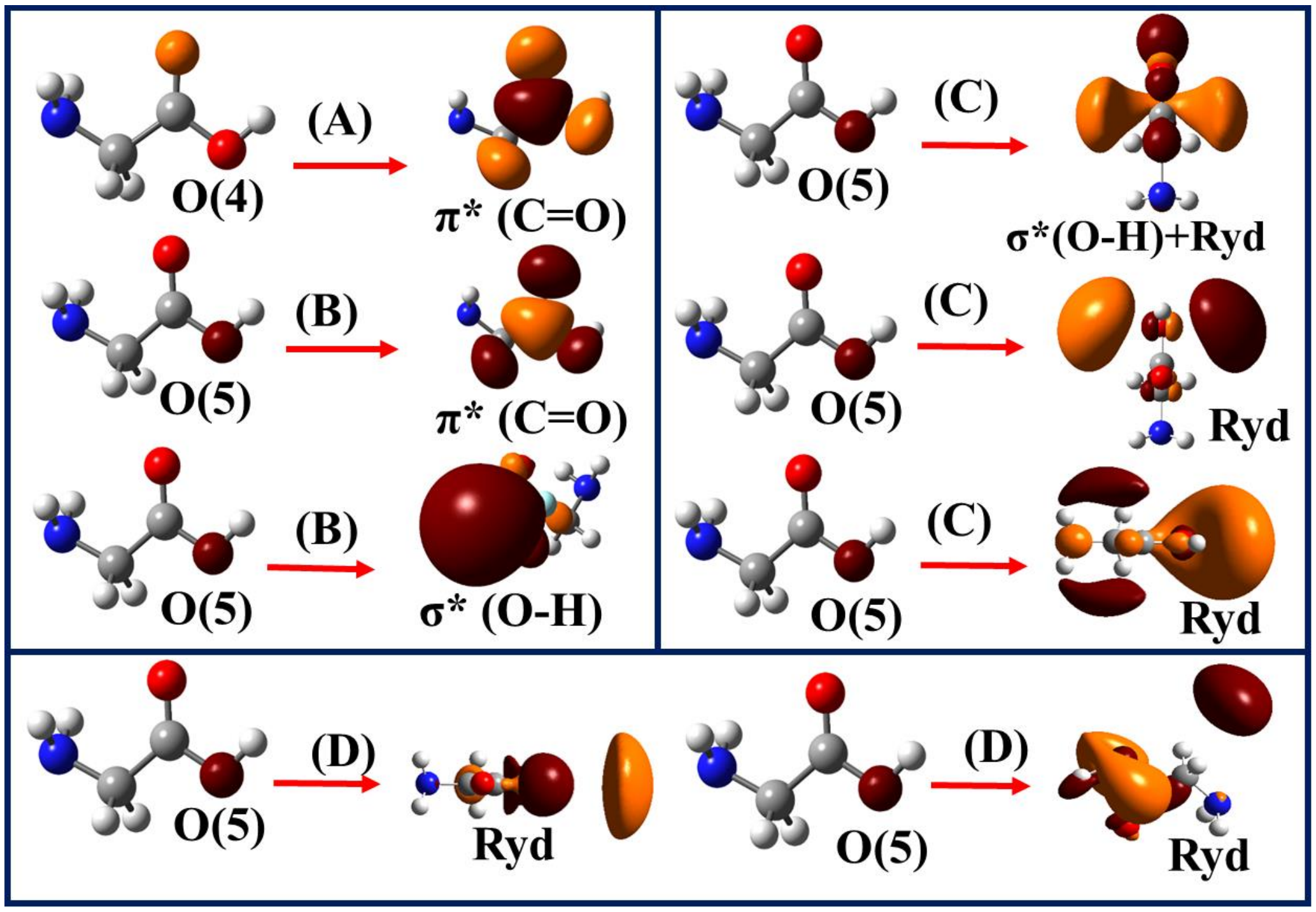

*Figure 8: Systematic representation of natural transition orbitals (NTOs) of O K edge of the glycine molecule.*

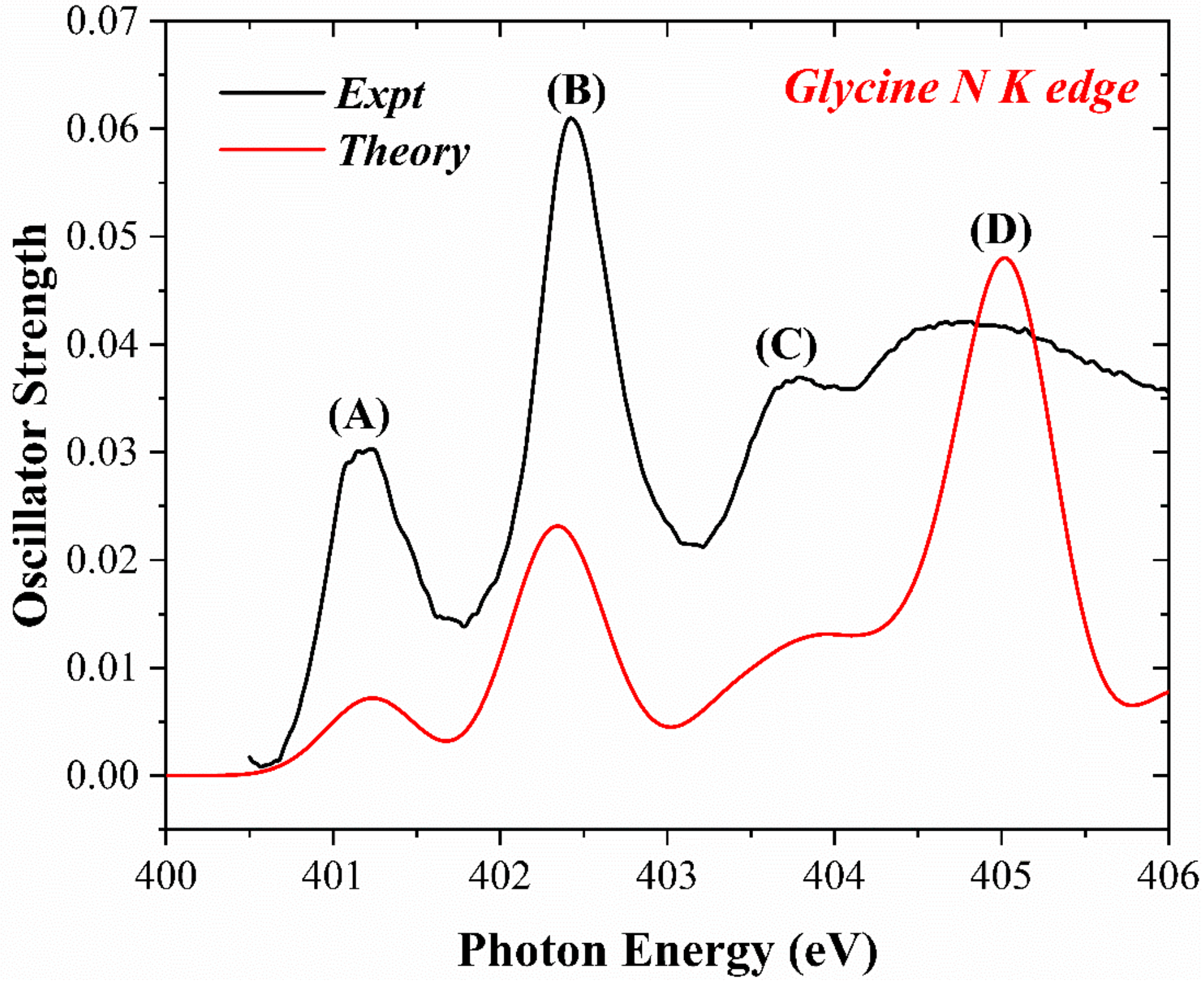

*Figure 9: Theoretical x-ray absorption spectra of N K-edge of glycine simulated at STEOM/aug-cc-pCVDZ level and it has been shifted by -3.04 eV to match with the experimental spectrum.*

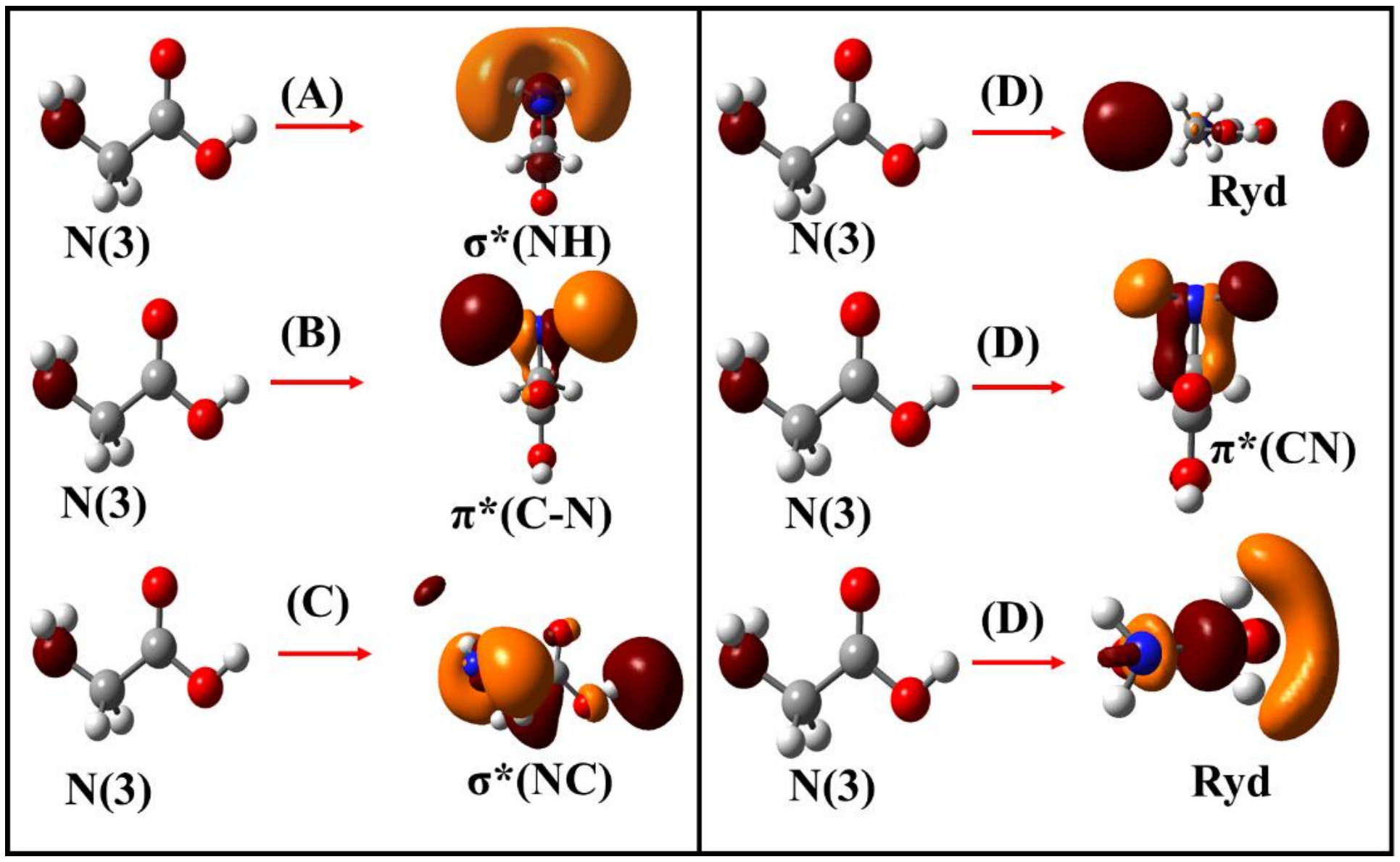

*Figure 10: Natural transition orbitals (NTOs) for the N K edge spectrum of the glycine molecule.*

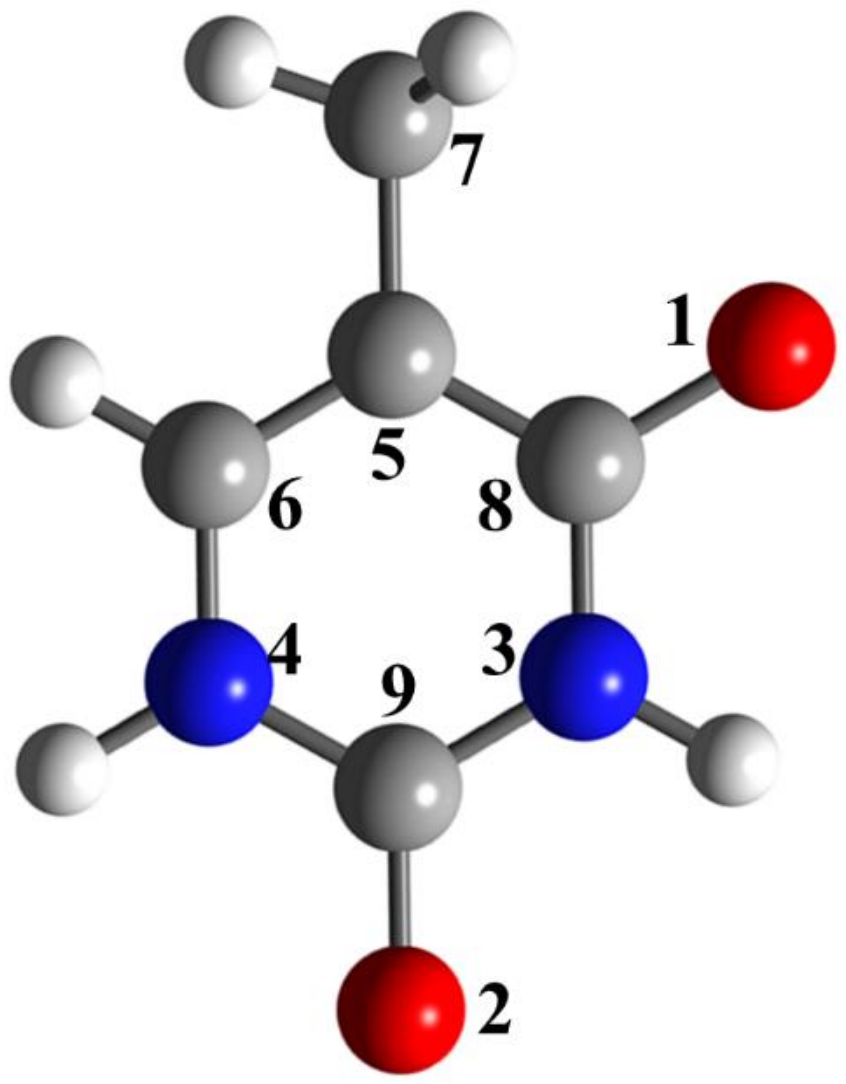

*Figure 11: The systematic representation of thymine (Oxygen = red, Nitrogen = blue, Carbon = grey, Hydrogen = white).*

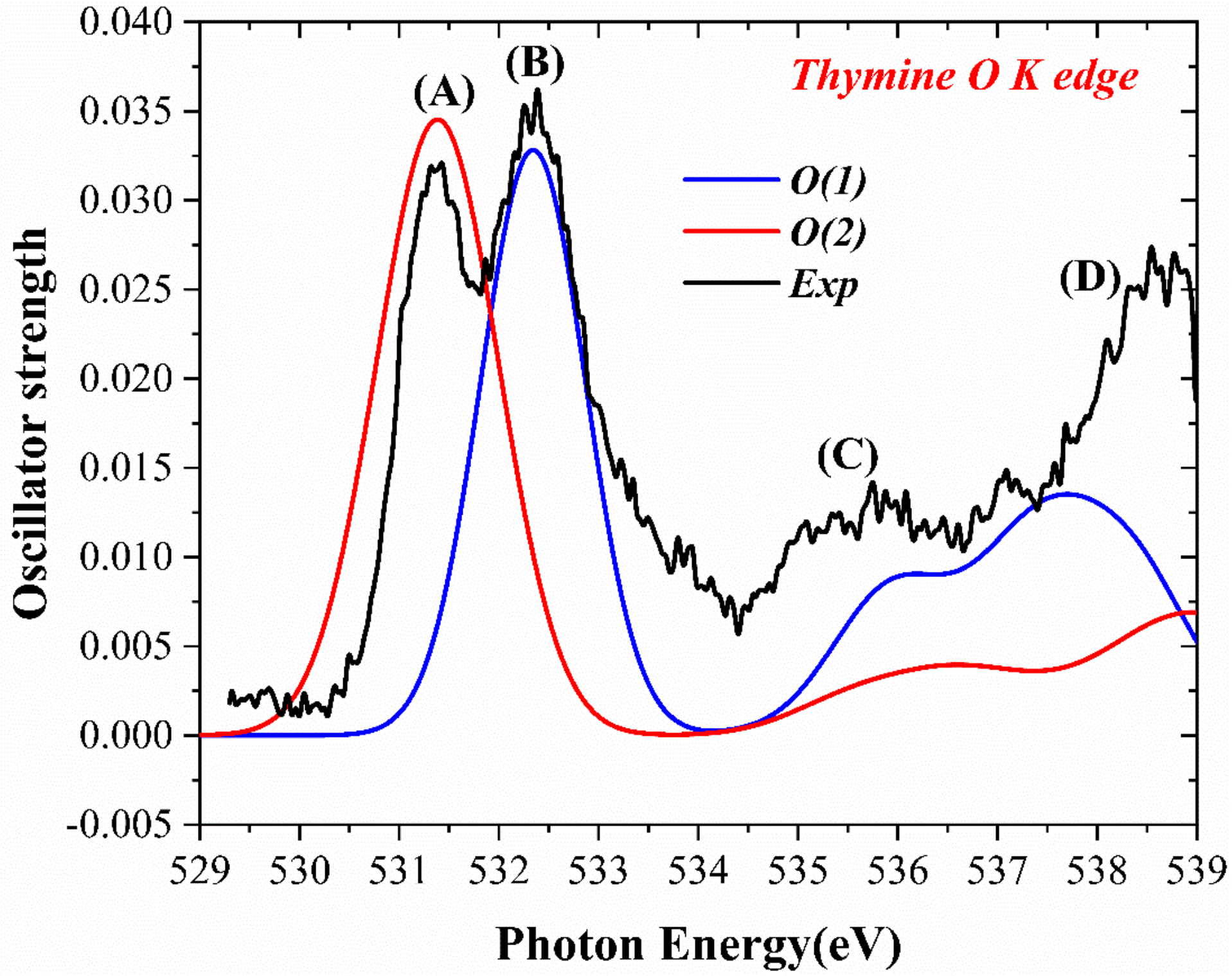

*Figure 12: Comparison of theoretical and experimental X-ray absorption spectra of oxygen K-edge in thymine. The simulated spectrum is shifted by -3.7 eV to align with the experimental spectrum.*

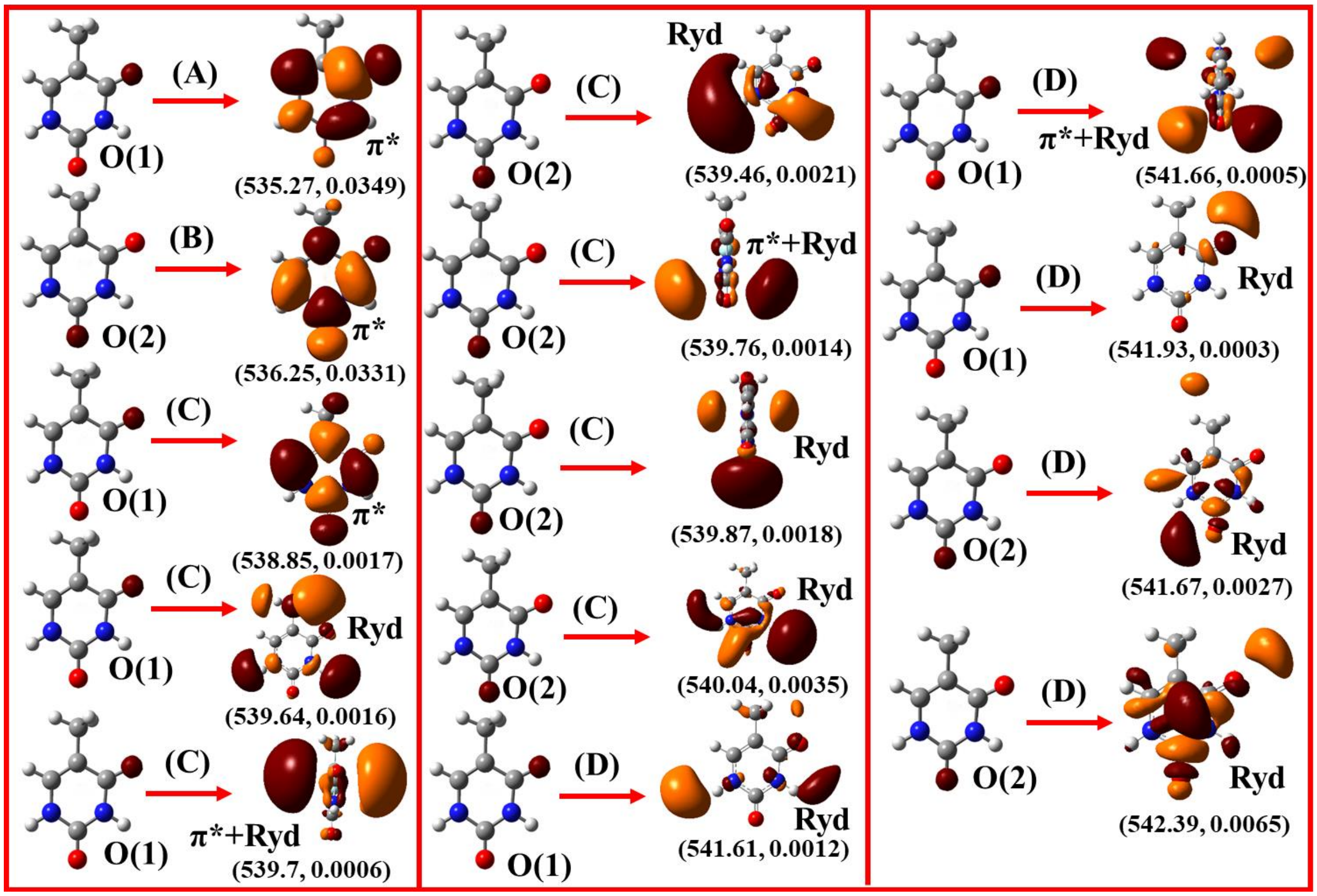

*Figure 13: Natural transition orbitals (NTO) for the oxygen K edge spectrum of thymine. All the core EE values mentioned are in eV and provided in the format (EE, Oscillator Strength).*

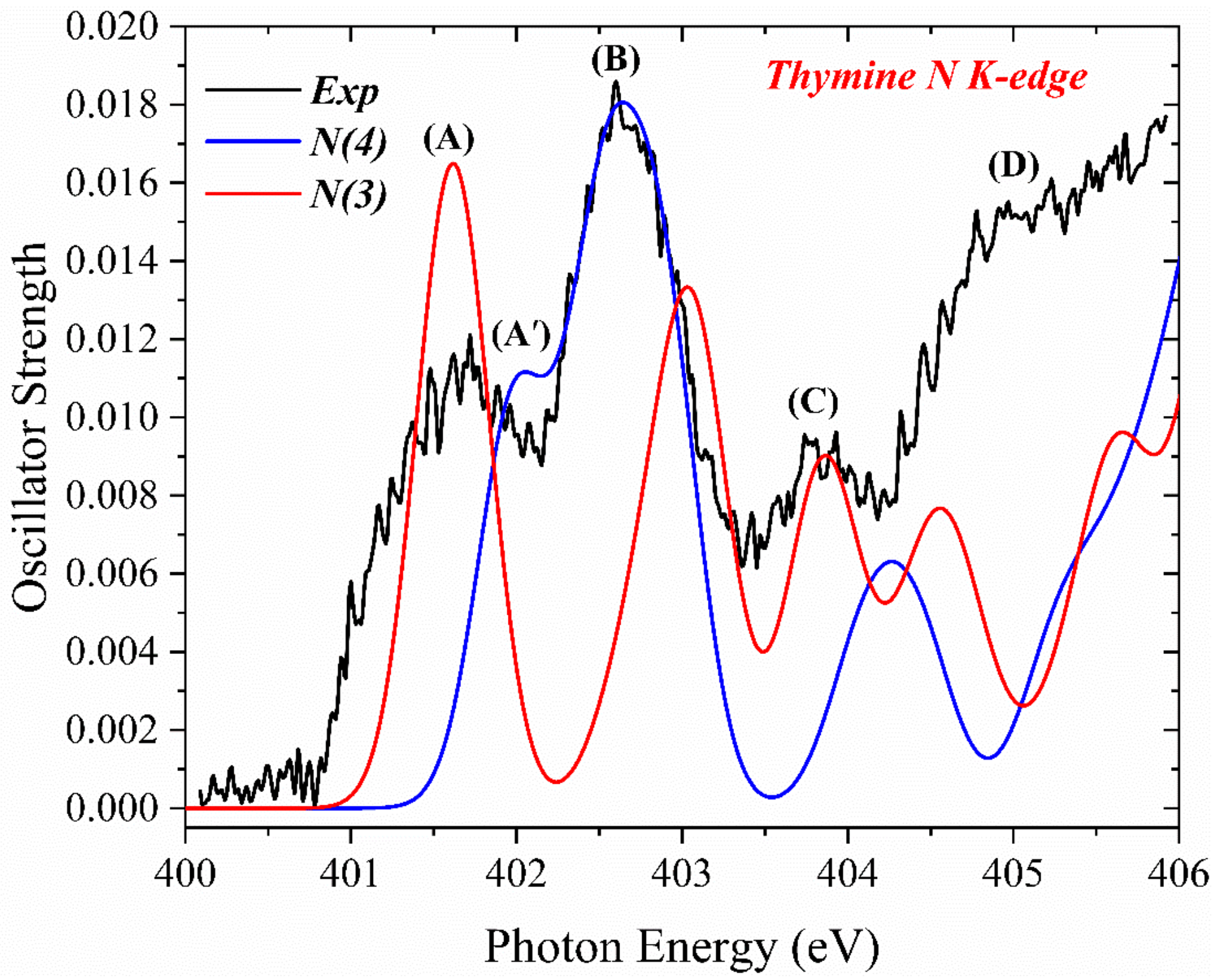

*Figure 14: Experimental and simulated nitrogen K-edge of thymine. The simulated spectra is shifted by -3.1 eV to match experimental spectrum.*

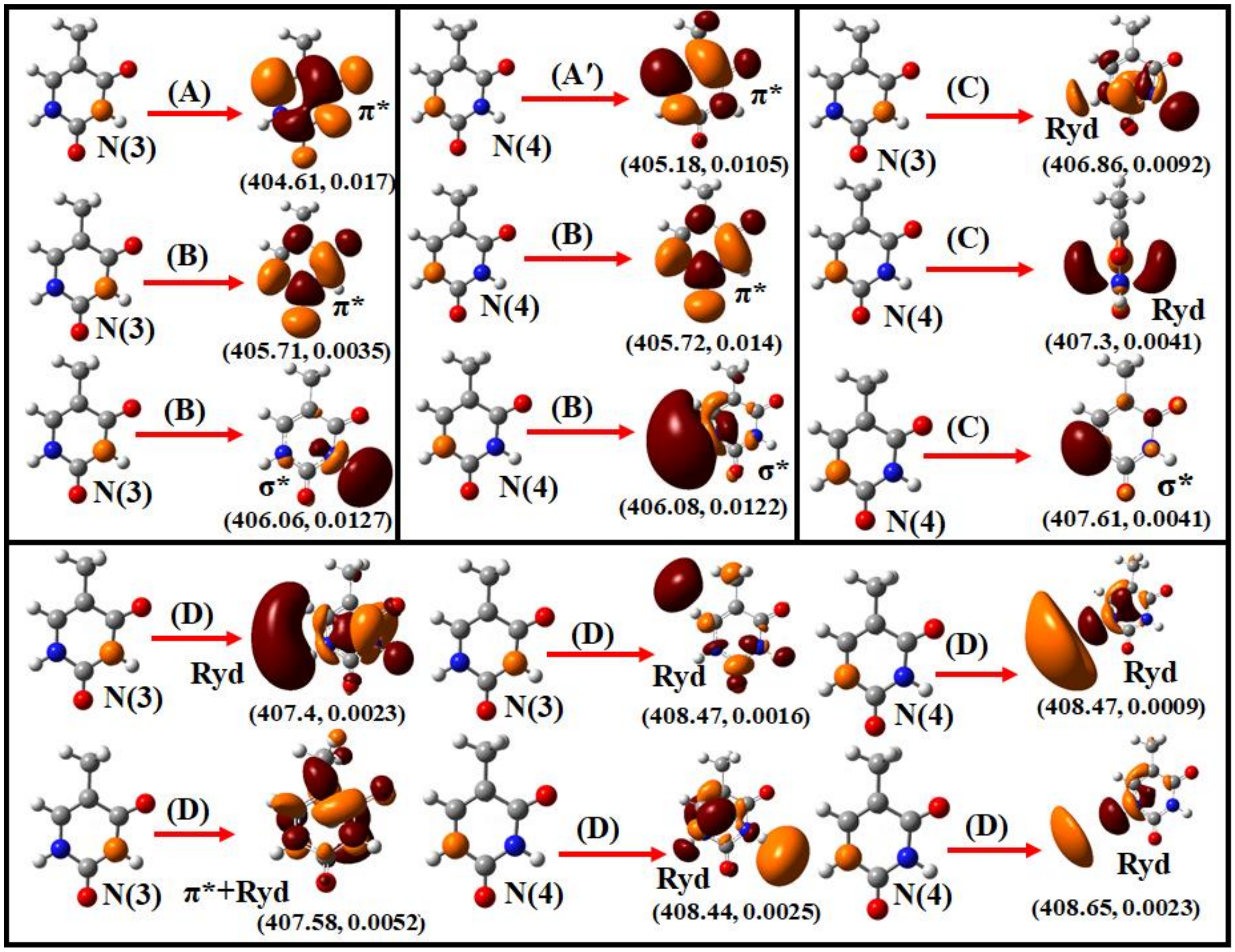

*Figure 15: Natural transition orbitals (NTOs) for the Nitrogen K edge spectrum of thymine. All the core EE values mentioned are in eV and provided in the format (EE, Oscillator Strength).*

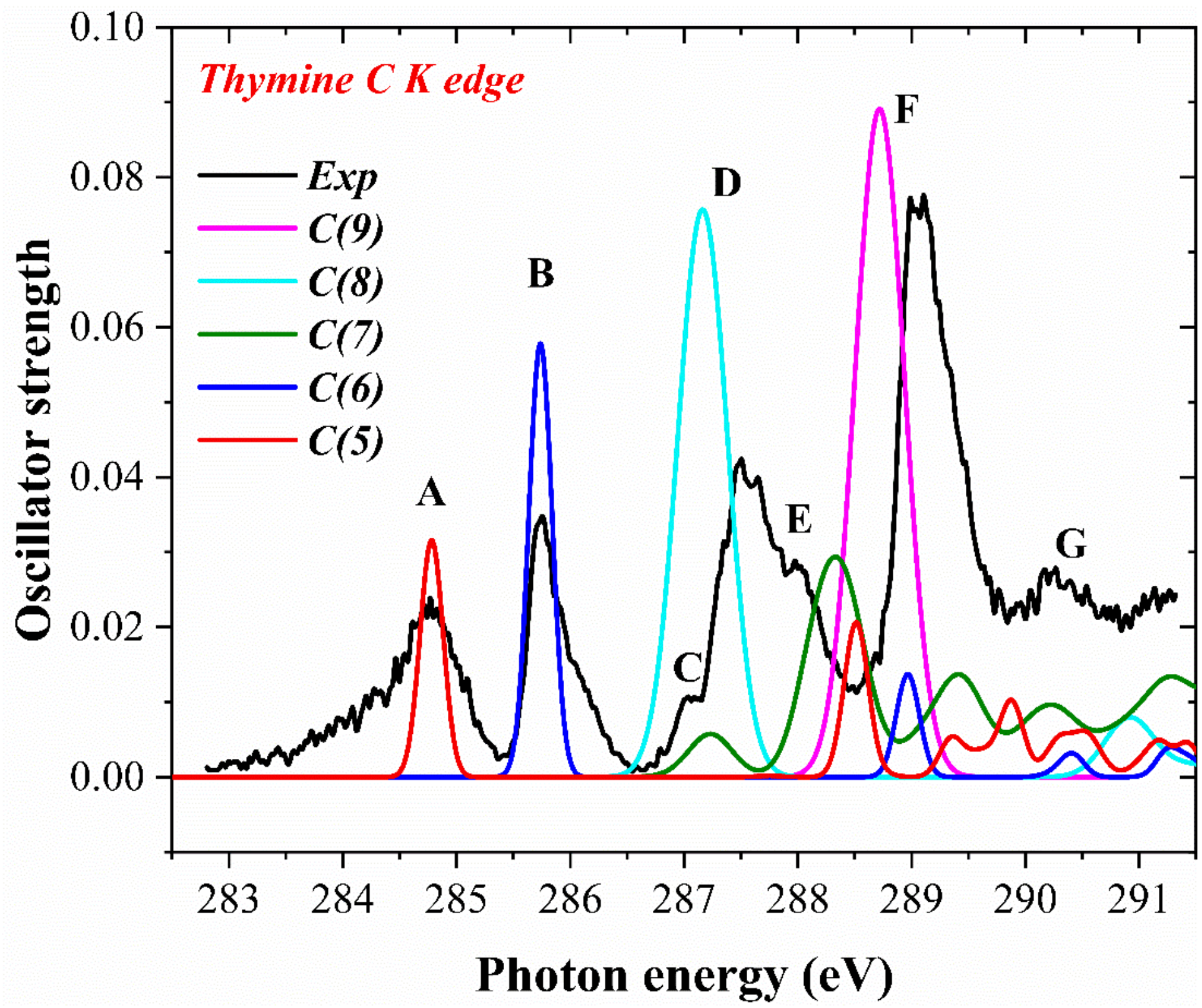

*Figure 16: Experimental and simulated carbon K-edge of thymine. The simulated spectra are shifted by -2.3 eV to match experimental spectrum.*

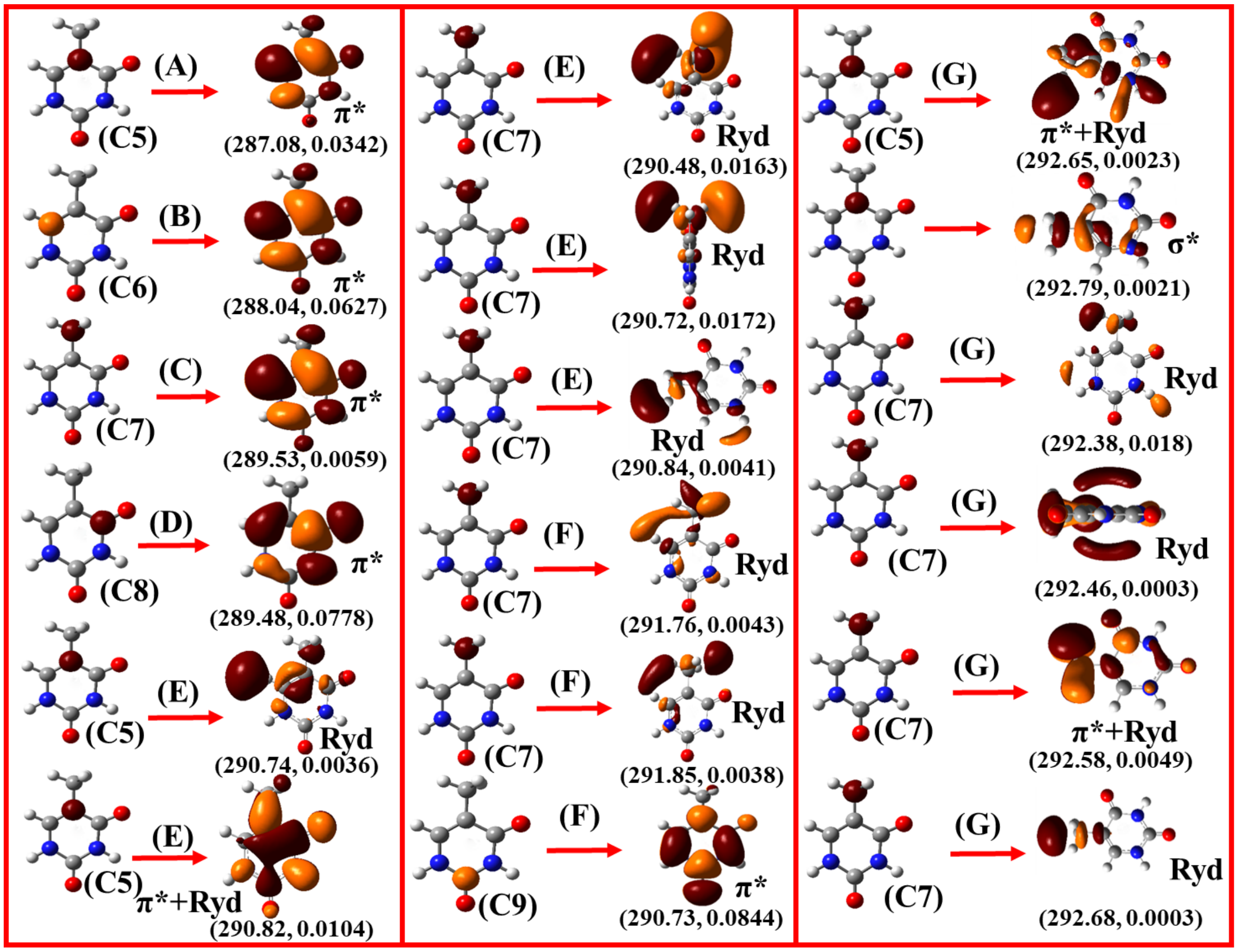

*Figure 17: Natural transition orbitals (NTOS) for the Carbon K edge spectrum of thymine. All the core EE values mentioned are in eV and provided in the format (EE, Oscillator Strength).*